\documentclass[preprint,review,12pt]{elsarticle}

\usepackage{amssymb}

\usepackage{amsmath,amsfonts}
\usepackage{algorithmic}
\usepackage{algorithm}
\usepackage{array}
\usepackage{textcomp}
\usepackage{stfloats}
\usepackage{url}
\usepackage{verbatim}
\usepackage{graphicx}
\usepackage{subfig}
\usepackage{float}
\usepackage{setspace}

\usepackage{bm}
\usepackage{multirow}
\usepackage{arydshln}
\usepackage{overpic}
\usepackage{color}
\usepackage{makecell}

\def\ie{\emph{i.e.}}
\def\eg{\emph{e.g.}}

\def\etal{{\em et al.}}

\newcommand{\cmark}{\ding{51}}

\graphicspath{{./Imgs/}{../../figure/}{./Imgs/results/}{./Imgs/authors/}}

\usepackage{times}
\usepackage{epsfig}
\usepackage{mathtools}
\usepackage{diagbox}
\usepackage{enumitem}
\usepackage{tablefootnote}
\newlist{todolist}{itemize}{2}
\setlist[todolist]{label=$\square$}
\usepackage{pifont}
\newcommand{\xmark}{\ding{55}}%
\newcommand{\done}{\rlap{$\square$}{\raisebox{2pt}{\large\hspace{1pt}\cmark}}\hspace{-2.5pt}}
\newcommand{\wontfix}{\rlap{$\square$}{\large\hspace{1pt}\xmark}}
\usepackage[pagebackref=true,breaklinks=true,colorlinks,bookmarks=false]{hyperref}

\usepackage{colortbl}
\usepackage{cleveref}
\crefformat{section}{\S#2#1#3} 
\crefformat{subsection}{\S#2#1#3}
\crefformat{subsubsection}{\S#2#1#3}

\usepackage{wrapfig} 
\usepackage{flushend} 

\journal{Pattern Recognition}

\begin{document}

\begin{frontmatter}

\title{FCNet: A Convolutional Neural Network for Arbitrary-Length Exposure Estimation}

\author[1,2]{Jin Liang}
\ead{nkujinliang@mail.nankai.edu.cn}

\author[1]{Yuchen Yang}
\ead{yycstat@mail.nankai.edu.cn}

\author[4]{Anran Zhang}
\ead{zhanganran@buaa.edu.cn}

\author[5]{Hui Li}
\ead{lihui@vivo.com}

\author[3]{Xiantong Zhen \corref{cor1}}
\ead{zhenxt@gmail.com}

\author[1]{Jun Xu}
\ead{nankaimathxujun@gmail.com}

\address[1]{School of Statistics and Data Science, Nankai University, Tianjin 300071, China}
\address[2]{School of Mathematics Science, Nankai University, Tianjin 300071, China}
\address[3]{Guangdong University of Petrochemical Technology, Maoming, Guangdong, China}
\address[4]{Key Laboratory of Advanced Technology of Near Space Information System, Beihang University, Beijing, China}
\address[5]{Imaging Algorithm Research Department, vivo Mobile Communication Co. Ltd., Shenzhen 518101, China}

\cortext[cor1]{Corresponding author}


\begin{abstract}
The photographs captured by digital cameras usually suffer from over or under exposure problems. For image exposure enhancement, the tasks of Single-Exposure Correction (SEC) and Multi-Exposure Fusion (MEF) are widely studied in the image processing community.
However, current SEC or MEF methods are developed under different motivations and thus ignore the internal correlation between SEC and MEF, making it difficult to process arbitrary-length sequences with improper exposures.
Besides, the MEF methods usually fail at estimating the exposure of a sequence containing only under-exposed or over-exposed images.
To alleviate these problems, in this paper, we develop a novel Fusion-Correction Network (FCNet) to tackle an arbitrary-length (including one) image sequence with improper exposures.
This is achieved by fusing and correcting an image sequence by Laplacian Pyramid (LP) image decomposition.
In each LP level, the low-frequency base component of the input image sequence is fed into a Fusion block and a Correction block sequentially for consecutive exposure estimation, implemented by alternative exposure fusion and correction.
The exposure-corrected image in current LP level is upsampled and fused with the high-frequency detail components of the input image sequence in the next LP level, to output the base component for the Fusion and Correction blocks in next LP level.
Experiments on the benchmark dataset~\cite{afifi2021learning} demonstrate that our FCNet is effective on arbitrary-length exposure estimation, including both SEC and MEF.
The code is publicly released at \url{https://github.com/NKUJinLiang/FCNet}.
\end{abstract}




\end{frontmatter}

\section{Introduction}
\label{sec:introduction}

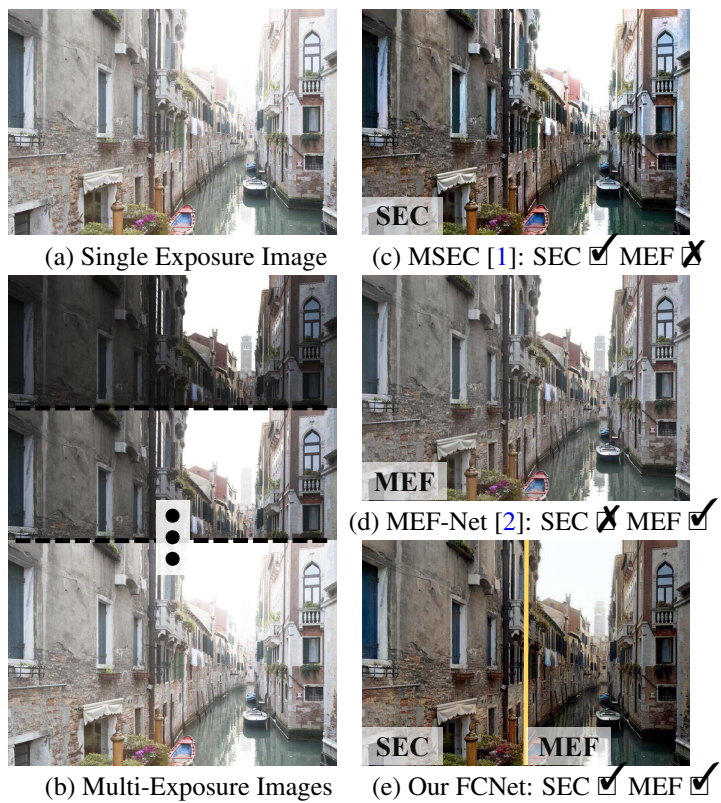
\begin{figure}[!t]
\begin{center}
\begin{overpic}[width=0.68\textwidth]{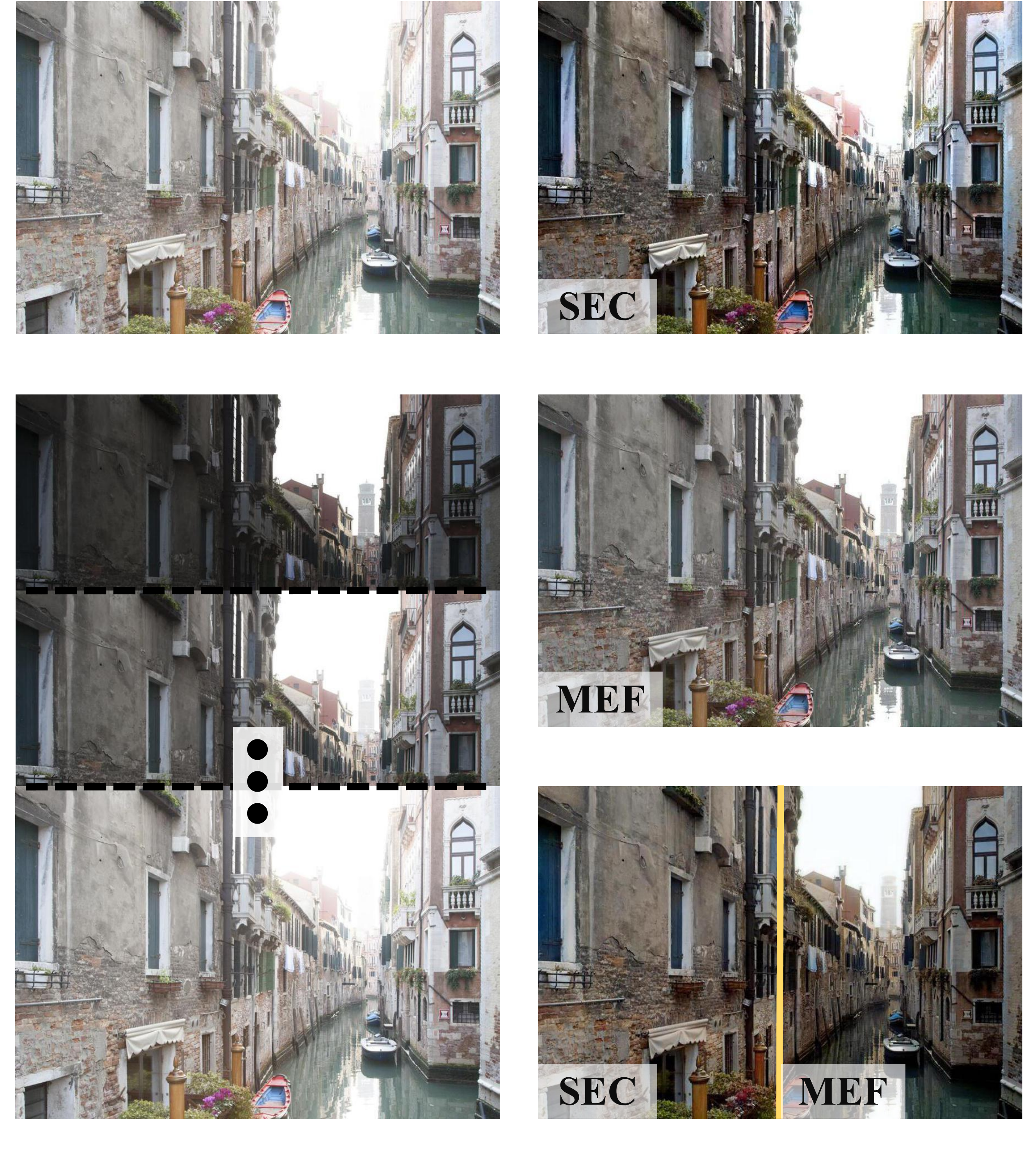}
\put(6,67){\footnotesize{(a) Single Exposure Image}}
\put(6,-1){\footnotesize{(b) Multi-Exposure Images}}
\put(48,67){\footnotesize{(c) MSEC~\cite{afifi2021learning}: SEC \done \ MEF \wontfix}}
\put(45,33){\footnotesize{(d) MEF-Net~\cite{ma2019deep}: SEC \wontfix \ MEF \done}}
\put(48,-1){\footnotesize{(e) Our FCNet: SEC \done \ MEF \done}}
\end{overpic}
\end{center}
\vspace{-1mm}
\caption{
\textbf{Enhanced images by SEC method~\cite{afifi2021learning}, MEF method~\cite{ma2019deep}, and the proposed FCNet}.
On one hand, the SEC method~\cite{afifi2021learning} well corrects the over-exposed image (a) and outputs the image (c), but cannot directly fuse the image sequence (b).
On the other, the MEF method~\cite{ma2019deep} well fuses the image sequence (b) and outputs the image (d), but cannot correct the image (a).
By integrating SEC and MEF in a unified framework, according to the input image (a) or sequence (b), our FCNet can output visual pleasing results (e) for SEC (a) (left side) and MEF (b) (right side).
}
\vspace{-6mm}
\label{figure:introduction}
\end{figure}

Exposure is an important aspect in camera to capture a photograph with proper brightness and visual quality~\cite{abdullah2007dynamic}.
During the capture process, exposure error is prone to happen due to improper shutter speed, focal-aperture ratio, or ISO value~\cite{afifi2021learning}.
This will degrade the visual quality of the captured single image or image sequence.
To address this issue, a number of methods have been proposed during the past decades, which can be roughly divided into the two categories: single-exposure correction methods~\cite{Liang_2018_CVPR,afifi2021learning} and multi-exposure fusion methods~\cite{ma2019deep,mertens2009exposure}.

Single-exposure correction (SEC) aims to generate a well-exposed and visually pleasing image from an under-exposed or over-exposed photograph~\cite{abdullah2007dynamic,afifi2021learning,Cai2018deep,jha2022camera}.
In the early stage, the under-exposure correction task is mainly tackled under the Retinex framework~\cite{land1977retinex}.
Most of existing Retinex-based methods~\cite{fu2016weighted,guo2016lime,JIA2023109823} mainly decompose an under-exposed image into the reflectance and illumination components, and then enhance the illumination one by gamma correction techniques~\cite{STAR2020}.
Though with promising performance, these Retinex-based methods would fail at handling over-exposed images.
%
Besides, the optimization process is usually time-consuming.
Recent SEC methods~\cite{afifi2021learning,Cai2018deep} employ deep neural networks to learn exposure correction from pairs of over/under-exposed and properly-exposed images.
However, these methods cannot be directly employed to perform exposure fusion, as they usually produce artifacts in challenging scenarios, which could not be properly presented in a single exposure image~\cite{mertens2009exposure}.
Generative SEC methods~\cite{jiang2021enlightengan} can recover the details in over-exposed areas to some extent, but often produce unnatural image colors.
Multi-exposure fusion (MEF) methods~\cite{ma2017robust,mertens2009exposure,YAN201966} generate visually appealing images by exploiting the complementary information from a sequence of over-exposed and under-exposed images.
These MEF methods integrate a series of images in low dynamic range with different exposures at the same scene to output a high-quality image~\cite{mertens2009exposure,Liang_2018_CVPR}, which usually suffer from high computational complexity upon a long sequence of high-resolution images.
To reduce the computational costs, several MEF methods~\cite{hasinoff2016burst,ma2019deep} implement computation exhaustive operations, \eg, alignment or mask prediction, on image sequences with downsampled resolutions.
However, this would produce unpleasant color bias or ghost artifacts in practical scenarios.
Besides, most MEF methods ignore the challenging scenarios with only under-exposed or over-exposed images, since only performing exposure fusion on these challenging cases is not enough to estimate a well-exposed image.
%
%
To this end, it is necessary to exploit the inner correlation between SEC and MEF for unified exposure fusion and correction framework.
In this paper, we propose a Fusion-Correction Network (FCNet) to perform exposure fusion and enhancement on arbitrary-length image sequences, naturally tackling the SEC and MEF tasks.
In our FCNet, we employ the Laplacian Pyramid decomposition~\cite{burt1987laplacian} to fuse and correct images alternately at multi-scale levels.
As shown in Figure~\ref{figure:introduction}, the SEC method~\cite{afifi2021learning} cannot directly tackle MEF, while the MEF method~\cite{ma2019deep} cannot directly tackle SEC.
By incorporating the fusion and correction capability into a unified network, our FCNet well handles both the SEC and MEF tasks. 

In summary, the contributions of our work are three-fold:
\begin{itemize}
\vspace{-1mm}
\item We propose a new Fusion-Correction Network (FCnet) to perform visual-pleasing exposure estimation on arbitrary-length image sequences.
Our FCNet can simultaneously tackle single-exposure correction (SEC) and multi-exposure fusion (MEF) tasks.

\item Under the Laplacian pyramid decomposition framework, our FCNet achieves efficient inference compared to the other competitors on handling long sequences of high-resolution images.

   \item Experiments show that our FCNet achieves comparable performance with existing methods on SEC, while better performance on MEF tasks, especially with over-exposed or under-exposed image sequences.
\end{itemize}

The rest of this paper is organized as follows.
In \S\ref{sec:related}, we briefly introduce the related works of our FCNet.
In \S\ref{sec:method}, we present our FCNet for SEC and MEF tasks.
Extensive experiments are conducted in \S\ref{sec:experiment} to evaluate the performance of our FCNet, with in-depth analysis.
\S\ref{sec:conclusion} concludes this work.

\begin{figure*}
\vspace{-2mm}
\begin{center}
\includegraphics[width=1\textwidth]{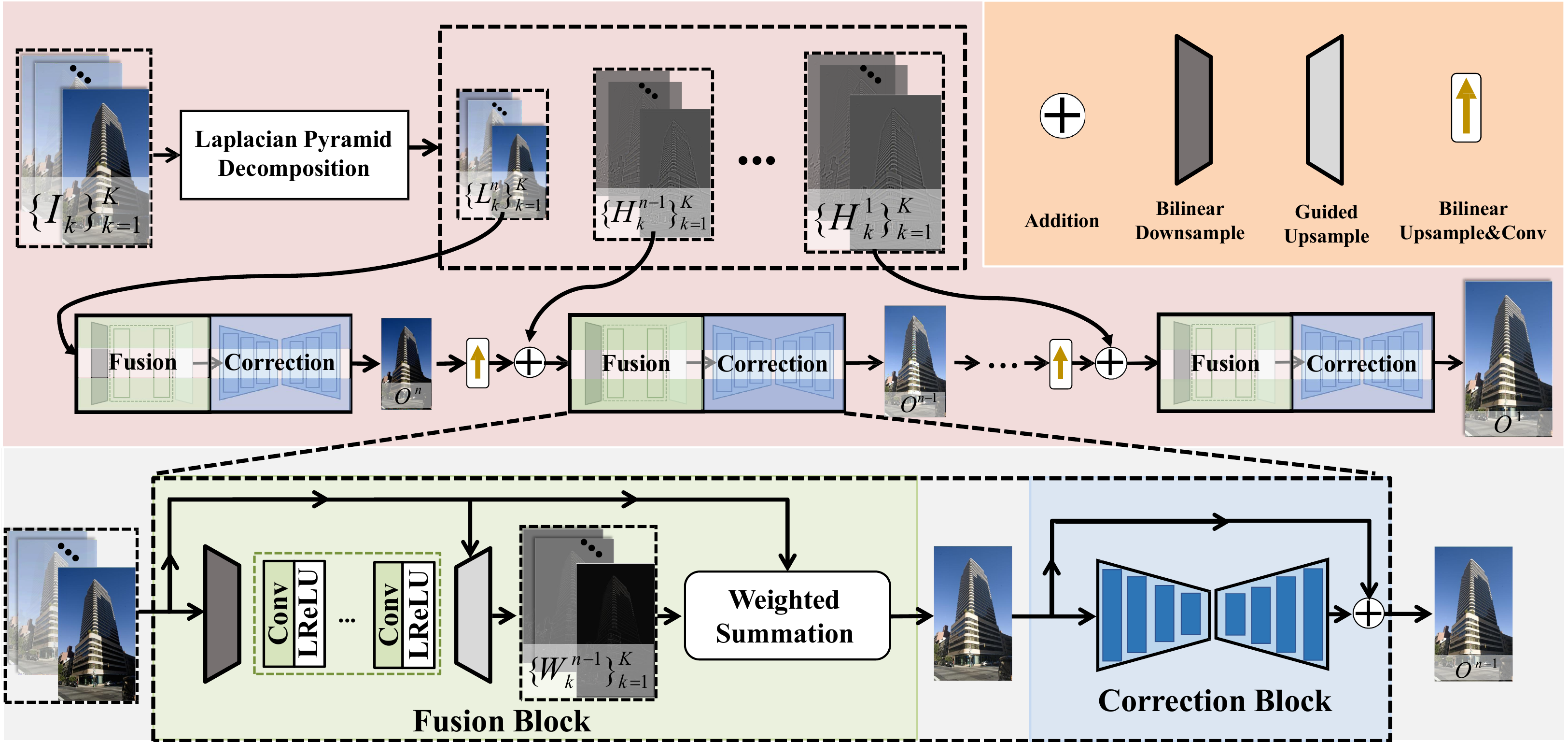}
\end{center}
\vspace{-4mm}
\caption{\textbf{Overview of the proposed FCNet for both single exposure correction and multi-exposure fusion}.
Given an input image sequence of arbitrary length $\mathcal{I}=\{{\mathbf{I}_k}\in\mathbb{R}^{h\times w}\}_{k=1}^{K}$ ($K\ge1$), we decompose the images into different Laplacian pyramid levels, and perform fusion and correction level-by-level in a coarse-to-fine manner.
Our FCNet sequentially fuses and corrects the  Laplacian pyramid images in each level, by a series of fusion and correction blocks.
The fusion block, in a light-weight structure, produces weigh maps for weighted image summation.
The correction block is in a UNet-like structure.
}
\vspace{-5mm}
\label{fig:short}
\label{fig:framework}
\end{figure*}

\section{Related Work}
\label{sec:related}
\noindent
\textbf{Single-exposure correction} (SEC) aims to improve the visual contrast of an under- or over-exposure image~\cite{}. The Retinex theory~\cite{land1977retinex} is widely studied to tackle the under-exposure problem by decomposing an image into the illumination and reflection components and correcting the illumination~\cite{fu2016weighted,guo2016lime,STAR2020,JIA2023109823}.
Fu \etal\cite{fu2016weighted} analysed the side effect of logarithmic transformation and designed a weighted variation model to refine the regularization terms.
In~\cite{guo2016lime}, Guo et al. first estimated a coarse illumination map and refined it by optimization techniques.
However, the Retinex-based methods tend to correct the illumination components in a consistent way, which may produce unnatural results when processing non-uniform illumination images. In addition, these methods have to solve complex optimization problems, which are often time-consuming.
Recently, deep learning methods~\cite{LI201815,chen2018learning,guo2020zero,jiang2021enlightengan,WANG2022108867,ZHOU2023109602} have achieved efficient performance on single image exposure correction.
Chen \etal\cite{chen2018learning} learned to directly enhance the under-exposed raw sensor images.
Guo \etal\cite{guo2020zero} proposed to estimate a series of adjustment curves for fast under-exposure image correction with carefully designed unsupervised loss functions. But this totally unreferenced training strategy also limits the generalization capability of the model, making it tend to generate over-exposed results in some cases. EnlightenGAN~\cite{jiang2021enlightengan} is a GAN-based network trained with unpaired under-exposed and well-exposed images. The adversarial loss enables the model to be trained in a wider range of image domains, especially the real-world scenarios in which paired images are hard to obtain.

However, the above methods mainly tackle under-exposed images, but often fail on over-exposure images which are also common in photography. The work of~\cite{afifi2021learning} is among the first deep learning based methods for both over- and under-exposure correction. Due to the limited exposure information in a single image, \cite{afifi2021learning} is not robust enough upon challenging scenarios. In this work, our FCNet can flexibly handle over- and under-exposure image correction tasks in an unified framework. In addition, with a multi-exposure image sequence, our FCNet restores the details of over- and under-exposed areas by integrating the information of different images.

\vspace{0mm}
\noindent
\textbf{Multi-exposure fusion} (MEF) is a promising alternative for high dynamic range imaging~\cite{mertens2009exposure,shen2014exposure}.
Early MEF methods mainly resort to pixel-wise operations.
Mertens \etal\cite{mertens2009exposure} proposed a multi-scale fusion mechanism via Laplacian pyramid decomposition~\cite{burt1987laplacian}, which is boosted by Shen \etal\cite{shen2014exposure} on computational efficiency.
To well preserve details in both bright and dark regions within a limited dynamic range, the methods of~\cite{bavirisetti2019multiguided,raman2009bilateral} separately process the base and detail components of an image decomposed by edge-preserving filtering~\cite{he2012guided,tomasi1998bilateral}. Gradient information is also exploited in~\cite{bertalmio2012variational,gu2012gradient,paul2016multi} for detail-enhanced MEF. However, these methods are not robust to the image sequences with misalignment~\cite{ma2017robust}.
For this, later methods~\cite{li2020fast,ma2017robust} usually perform patch-wise MEF to optimize the MEF-SSIM metric~\cite{ma2015perceptual}.
But patch-based methods are prone to produce over-smooth results upon dynamic image sequences.
Recently, deep MEF methods~\cite{ma2019deep,ram2017deepfuse,xu2020GAnmef} have achieved robust performance upon challenging scenarios.
DeepFuse~\cite{ram2017deepfuse} converted the image sequence to YCbCr format and performed fusion on the Y channel.
To achieve efficient MEF, Ma \etal\cite{ma2019deep} proposed to predict fusion maps in a downsampled resolution, and upsampled the maps to the original resolution for final fusion.
MEF-GAN~\cite{xu2020GAnmef} achieved visually appealing results using the generative adversarial networks~\cite{goodfellow2014generative}. Overall, these MEF methods could only perform image fusion and the luminance of the resulting image can only be within the luminance range of the input image sequence. In this paper, we explore an integrated framework to tackle both the SEC and MEF tasks. Due to the additional correction ability, our FCNet can handle more challenging scenes, such as fusing and correcting image sequence containing only under-exposed or over-exposed images.
%

\begin{figure}[!t]
\centering
\captionsetup[subfloat]{labelsep=none,format=plain,labelformat=empty,font={scriptsize,stretch=0.3}}
\subfloat[(a) Input Base Sequence]{
    \includegraphics[scale=0.38]{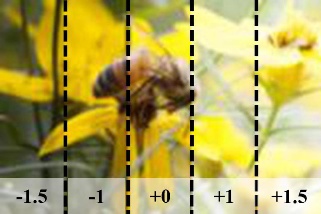}}
\subfloat[(b) level 4 ($\mathbf{O}^{4}$)]{
    \includegraphics[scale=0.38]{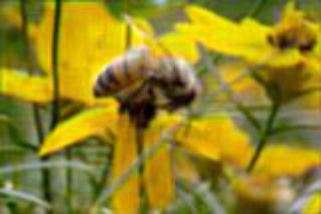}}
\subfloat[(c) level 3 ($\mathbf{O}^{3}$)]{
    \includegraphics[scale=0.38]{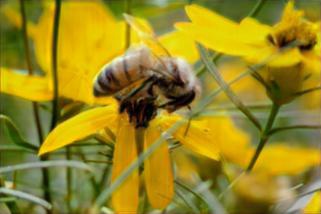}}
\\
\vspace{-0.2cm}
\subfloat[(d) level 2 ($\mathbf{O}^{2}$)]{
    \includegraphics[scale=0.38]{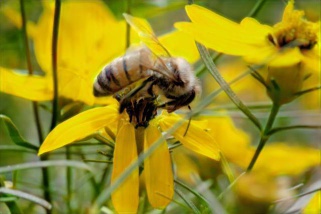}}
\subfloat[(e) level 1 ($\mathbf{O}^{1}$)]{
    \includegraphics[scale=0.38]{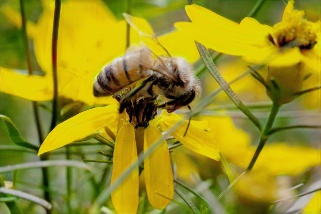}}
\subfloat[(f) Ground-Truth]{
    \includegraphics[scale=0.38]{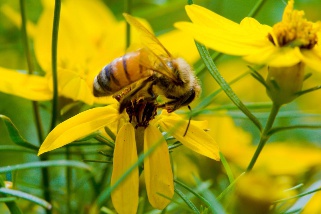}}
\vspace{-0.2cm}
\caption{\textbf{Intermediate output images $\mathbf{O}^{i}$ by the FC blocks of our FCNet in different Laplacian pyramid levels} on an image sequence (with EVs of -1.5, -1, +0, +1, +1.5).}
\vspace{-0.4cm}
\label{fig:process flow}
\end{figure}

\section{Proposed Fusion-Correction Network}
\label{sec:method}
Our Fusion-Correction Network (FCNet) mainly contains three parts: Laplacian Pyramid decomposition (\S\ref{sec:LP}), Fusion-Correction (FC) block (\S\ref{sec:FC}), and Base-Detail composition between two FC blocks (\S\ref{sec:Base-Detail}), as shown in Figure~\ref{fig:framework}.

\subsection{Laplacian Pyramid Decomposition}
\label{sec:LP}
%
To tackle SEC or MEF, our FCNet needs to extract multi-scale features of the input images and reconstruct multi-scale contents of the output image. And the image decomposition can be utilized to fulfill the multi-scale processing. In our FCNet, we employ the Laplacian Pyramid (LP) decomposition~\cite{burt1987laplacian} to perform exposure enhancement in a coarse-to-fine manner.
Given a sequence of $K\ge1$ images $\{ {\mathbf{I}_k}\in\mathbb{R}^{h\times w}\}_{k=1}^{K}$ of arbitrary exposures, we first decompose each image ${\mathbf{I}_k}$ into $n$ hierarchical layers, including $n-1$ high-frequency detail components ${\mathcal{H}_k} = \{ \mathbf{H}_k^1,\mathbf{H}_k^2,...,\mathbf{H}_k^{n - 1}\}$ and the low-frequency base component ${\mathbf{L}_k^n}$ at the $n$-{th} layer.
Here, $\mathbf{H}_k^i$ ($i=1,...,n-1$) is of size $\frac{1}{2^{i-1}}h\times\frac{1}{2^{i-1}}w$ while  ${\mathbf{L}_k^n}$ is of size $\frac{1}{2^{n-1}}h\times\frac{1}{2^{n-1}}w$.
The final image is reconstructed from these components sequentially by the proposed Fusion-Correction block introduced as follows.

\subsection{Fusion-Correction Block}\label{sec:FC}
\begin{figure}[t]
    \begin{center}
        \includegraphics[width=\linewidth]{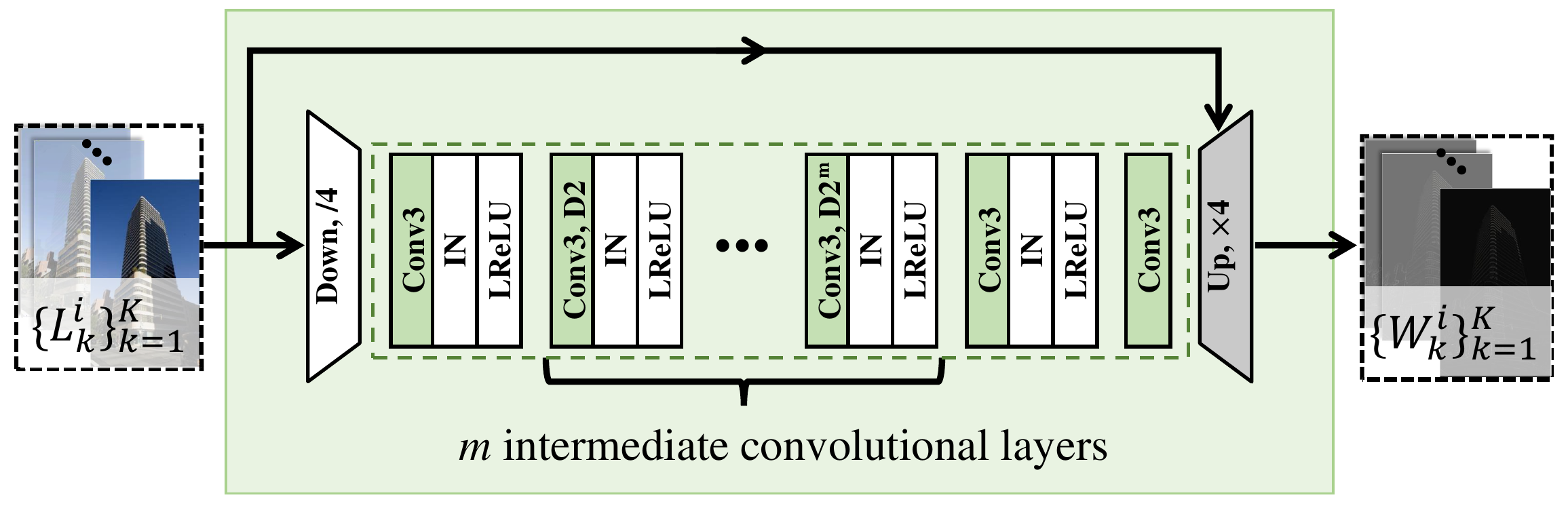}
    \end{center}
    \vspace{-0.6cm}
    \caption{\textbf{Detailed convolution layers architecture of the fusion block}.
    In the fusion block, we use ``Downsample-Execute-Upsample'' scheme~\cite{ma2019deep} to reduce the computational costs.
    ``Downsample, /4'' is bilinear downsampling, and ``Upsample, $\times$4'' is guided upsampling~\cite{he2012guided}.
    The intermediate convolutional layers are dilated convolutions.}
    \label{fig: fusion_block}
    \vspace{-0.4cm}
\end{figure}

To tackle SEC and MEF in an integrated framework, we propose a Fusion-Correction (FC) block to consecutively enhance the image sequence under the LP framework.
Given the $n$-th base image sequence $\{\mathbf{L}_k^{n}\}_{k=1}^K$, our FC block first performs a weighted sum of the base sequence by a Fusion block and then enhances the fusion result by a U-Net like Correction block.
The output image $\mathbf{O}^{n}$ will be upsampled and composed with the high-frequency components $\{\mathbf{H}_k^{n-1}\}_{k=1}^K$ at level $n-1$, to generate the base sequence $\{\mathbf{L}_k^{n-1}\}_{k=1}^K$ as the input of the next FC block.
Our FC block alternately performs on the base components $\{\mathbf{L}_k^{i}\}_{k=1}^K$ ($i=n,...,1$) and composition on ${\mathbf{O}^i}$ with $\{\mathbf{H}_k^{i-1}\}_{k=1}^K$ ($i=n,...,2$) to output the final image ${\mathbf{O}^1}$ (we do not perform composition when $i=1$).
\noindent
\textbf{Fusion block}.
We perform image fusion before exposure correction to avoid large memory consumption and computational costs, when dealing with a long image sequence.
Besides, a good fusion result tends to have rich details and little noise~\cite{twsc2018}, reducing the difficulty of exposure correction that is prone to bring unpleasing artifacts~\cite{Cai2018deep}.
Inspired by the acceleration schemes in~\cite{hasinoff2016burst,gharbi2017deep,ma2019deep}, we design our fusion block under the ``Downsample-Execute-Upsample'' scheme~\cite{ma2019deep} to reduce the computational costs of our model upon long image sequences.
Specifically, the Fusion block in the $i$-{th} ($i=1,...,n$) LP level first downsamples the base sequence $\{\mathbf{L}_k^i\}_{k=1}^{K}$ by a bilinear interpolation and predicts the weight maps $\{\mathbf{W}_k^i\}_{k=1}^{K}$ by several dilated convolutions~\cite{yu2015multi} at a downsampled size of $\{\mathbf{L}_k^i\}_{k=1}^{K}$.
Then we upsample the weight maps $\{\mathbf{W}_k^i\}_{k=1}^{K}$ with the guidance of the input sequence $\{\mathbf{L}_k^i\}_{k=1}^{K}$~\cite{he2012guided} to fuse them at their original resolutions.
The fused image $\mathbf{F}^i$ in the $i$-{th} layer is obtained as follows:
\begin{equation}
\setlength\abovedisplayskip{1pt}
\setlength\belowdisplayskip{1pt}
\begin{split}
\mathbf{F}^i = \sum\limits_{k = 1}^K {{g_\uparrow}(\mathbf{W}_k^i) \otimes \mathbf{L}_k^i},
\end{split}
\end{equation}
where $\otimes$ is the Hadamard product and $g_\uparrow$ denotes the guided upsampling operation~\cite{he2012guided}.
The dilated convolutions used in each Fusion block are shown in Figure~\ref{fig: fusion_block}.
In each Fusion block, we use $m$ intermediate convolutional layers between the first convolutional layer and last two convolutional layers.
Here, we set $m=3,2,1,0$ in corresponding LP levels.
Thus, the size of Fusion block decreases gradually in different LP levels.

\begin{figure}[t]
    \begin{center}
        \includegraphics[width=\linewidth]{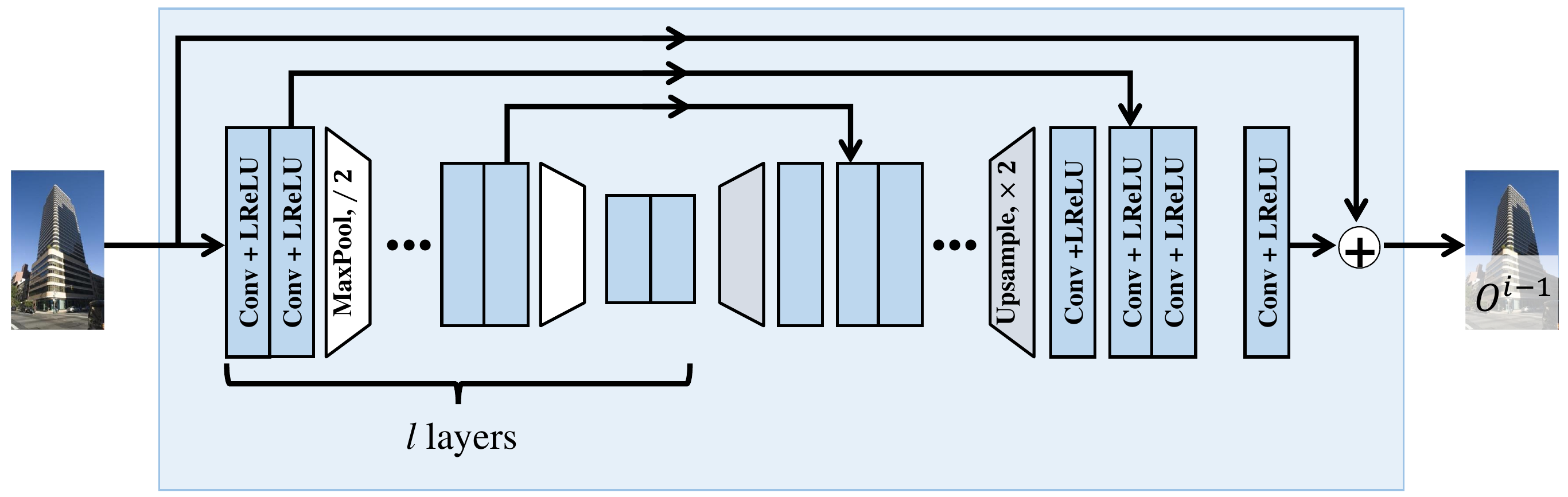}
    \end{center}
    \vspace{-0.6cm}
    \caption{\textbf{Detailed convolution layers architecture of the correction block}.
    We use a UNet-like architecture for the correction block.
    ``Upsample, $\times$2'' is bilinear upsampling.}
    \label{fig: correction_block}
    \vspace{-0.4cm}
\end{figure}

\noindent
\textbf{Correction block}.
Given the fused image $\mathbf{F^i}$, a Correction block at the $i$-th layer is applied on it to produce an exposure-enhanced image $\mathbf{O}^{i}$.
Since the UNet~\cite{ronneberger2015u} proves to be effective in many low-level tasks~\cite{eilertsen2017hdr,liu2020single,hu2019runet}, our Correction block is also designed as a UNet-like network.
To alleviate potential checkerboard artifacts~\cite{jiang2021enlightengan}, we replace the standard deconvolutional layer at each upsampling stage of UNet by a bilinear upsampling layer with a $3\times3$ convolutional layer.
To improve its correction capability, we add a skip connection between the encoder and decoder of the UNet-like network.
As a standard UNet, the number of output channels in the encoder is doubled from layer to layer with halved image resolution, as shown in Figure~\ref{fig: correction_block}.
The Correction block in the base LP layer is a 4-layer UNet-like network, which has 24 channels in the first convolutional layer of the encoder.
The second Correction block is also a 4-layer UNet-like network and has 16 channels in the first convolutional layer of the encoder.
The third and the final Correction blocks are 3-layer UNet-like networks with 16 channels in the first convolutional layer of the encoder.
Overall, the size of FC block gradually decreases with the progress of the image reconstruction.
This enables our FCNet to focus more on fusing and correcting the base LP levels of the input image or image sequence.
The ablation studies in~\S\ref{sec:ablation} validate the effectiveness of this structure design for our Fusion and Correction blocks.

\subsection{Base-Detail Composition}
\label{sec:Base-Detail}
The composition between two FC blocks in FCNet is exactly the composition operator of Laplacian Pyramid decomposition~\cite{burt1987laplacian}.
The overall motivation of FCNet is to fuse the image fusion module and correction module into Laplacian Pyramid decomposition and composition to handle both Single-Exposure Correction (SEC) and Multi-Exposure Fusion (MEF) tasks in an integrated framework.
As the image reconstruction process under the LP framework~\cite{burt1987laplacian}, the image $\mathbf{O}^i$ output by the FC block at layer $i$ ($i=n,...,2$) will be upsampled and composed with the high-frequency components $\{\mathbf{H}_k^{i-1}\}$ in layer $i-1$.
The resulting base sequence $\{\mathbf{L}_k^{i-1}\}$ is the input of the FC block at layer $i-1$.
The composition process is formulated as:
\begin{equation}
\setlength\abovedisplayskip{1pt}
\setlength\belowdisplayskip{1pt}
\begin{split}
\mathbf{L}_k^{i-1} = {f_\uparrow}({\mathbf{O}^i}) + \mathbf{H}_k^{i-1}, k=1,...,K,
\end{split}
\end{equation}
where $f_\uparrow$ denotes the upsampling operation implemented by a bilinear interpolation layer and a $3\times3$ convolutional layer. The intermediate images output in different stages of our FCNet are shown in Figure~\ref{fig:process flow}. One can see that our FCNet successively enhances the input image sequence at different levels.

\subsection{FCNet for Image Sequence at Arbitrary Length}
\label{sec:fusion of arbitrary length}
The proposed FCNet is able to process an image sequence of arbitrary lengths (including one) by an integrated convolution neural network.
For an input image sequence of arbitrary length, our FCNet concatenates the images along the batch dimension.
And each image is fed into the fusion block to predict a corresponding weight map. The weight maps are obtained by a parallel way as same as the regular batch operation.
When using the weight maps to fuse an image sequence, we normalize the weight maps by dividing the sum of the weight maps along the batch dimension.
Based on this design, the fusion module in our FCNet can fix the number of input channels as 3 and generate a series of 3-channel weight maps according to the number of input sequences for image fusion.
In the training stage, for each image sequence, we select a random number of images as the inputs.
During the inference stage, a single trained FCNet can process an image sequence of arbitrary lengths in the same way.

\subsection{Loss Function}
\label{sec:loss}
To endow our FCNet with the capability to well handle the SEC and MEF tasks simultaneously, our FCNet is trained end-to-end by the following loss function:
\begin{equation}\label{loss:reconstruction}
\setlength\abovedisplayskip{1pt}
\setlength\belowdisplayskip{1pt}
\mathcal{L} = {\mathcal{L}_{pr}} + {\lambda }{\mathcal{L}_{ps}},
\end{equation}
where $\mathcal{L}_{pr}$ is the pyramid reconstruction loss, ${\mathcal{L}_{ps}}$ is the pyramid spatial consistency loss, and $\lambda$ is the weight used to trade-off different terms.

\begin{figure}[!t]
\centering
\captionsetup[subfloat]{labelsep=none,format=plain,labelformat=empty,font={scriptsize,stretch=0.3}}
\subfloat[Input Sequence]{
    \includegraphics[scale=0.141]{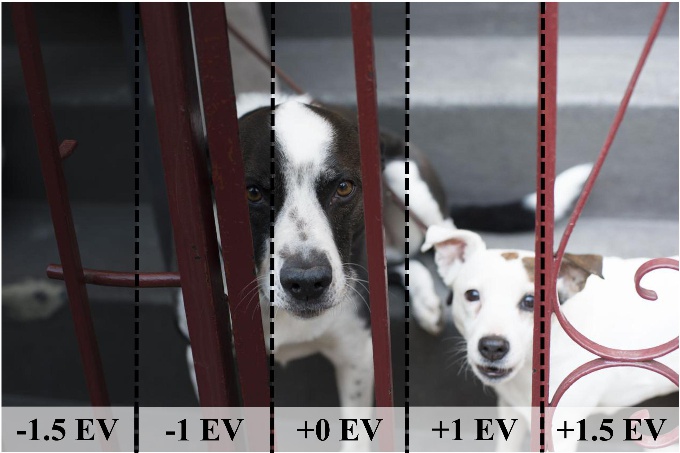}}
\subfloat[Mertens09~\cite{mertens2009exposure}]{
    \includegraphics[scale=0.30]{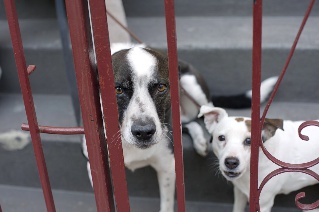}}
\subfloat[MEF-GAN~\cite{xu2020GAnmef}]{
    \includegraphics[scale=0.30]{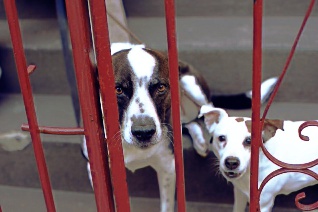}}
\subfloat[MEF-Net~\cite{ma2019deep}]{
    \includegraphics[scale=0.30]{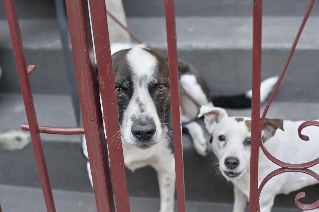}}
\\
\vspace{-0.2cm}
\subfloat[Zero-DCE~\cite{guo2020zero}]{
    \includegraphics[scale=0.30]{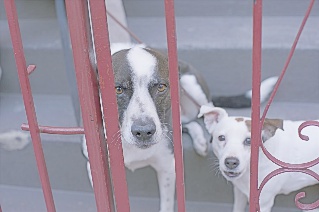}}
\subfloat[MSEC~\cite{afifi2021learning}]{
    \includegraphics[scale=0.30]{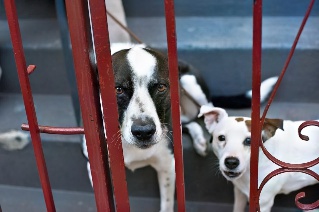}}
\subfloat[FCNet (Ours)]{
    \includegraphics[scale=0.30]{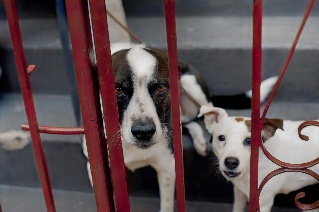}}
\subfloat[Ground-Truth]{
    \includegraphics[scale=0.30]{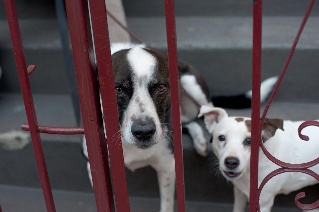}}
\vspace{-0.2cm}
\caption{\textbf{Qualitative comparison on a multi-exposure sequence} (with EVs of -1.5, -1, +0, +1,+1.5) by different MEF methods (Mertens09~\cite{mertens2009exposure}, MEF-GAN~\cite{xu2020GAnmef}, MEF-Net~\cite{ma2019deep}) and SEC methods (Zero-DCE~\cite{guo2020zero}, MSEC~\cite{afifi2021learning}).}
\vspace{-0.4cm}
\label{fig:MEF comparison results 1}
\end{figure}

\noindent
\textbf{Pyramid reconstruction loss}.
As suggested by~\cite{afifi2021learning}, we decompose the ground-truth image $\mathbf{G}$ into a Gaussian pyramid and use each pyramid layer to supervise the corresponding output by the FC block at that layer, as follows:
\begin{equation}
\setlength\abovedisplayskip{1pt}
\setlength\belowdisplayskip{1pt}
\label{loss:pr}
{\mathcal{L}_{pr}} = \sum\limits_{i = 2}^n {{2^{(i-2)}}} \sum\limits_{p = 1}^{3{h_i}{w_i}} {\left| {{\mathbf{O}^i}(p) - {\mathbf{G}^i}(p)} \right|}+\sum\limits_{p = 1}^{3{h}{w}} {\left| {\mathbf{O}^{1}(p) - \mathbf{G}(p)} \right|},
\end{equation}
where $\mathbf{O}^{i}$ is the output image of $i$-{th} FC block and $\mathbf{G}^{i}$ is the $i$-{th} pyramid layer decomposed from the ground truth image $\mathbf{G}$.
${h_i}={\frac{1}{2^{i-1}}h}$ and ${w_i}={\frac{1}{2^{i-1}}w}$ are the height and width of $\mathbf{O}^{i}$.
The pyramid reconstruction loss $\mathcal{L}_{pr}$ supervises the intermediate output of each FC block, making our FCNet robust to large exposure deviation during the reconstruction.
Our ablation study in \S\ref{sec:ablation} also validates the effectiveness of the pyramid reconstruction loss on exposure correction.

\noindent
\textbf{Pyramid spatial consistency loss}.
To improve our FCNet robustness upon challenging images with large homogeneous regions, here we propose a pyramid spatial consistency loss ${\mathcal{L}_{ps}}$ to preserve the difference of neighboring regions between the output image and its corresponding ground-truth.
Inspired by~\cite{guo2020zero}, we divide the image $\mathbf{O}^{i}$ (output by the FC block at layer $i$) and the $i$-th pyramid ground-truth layer $\mathbf{G}^{i}$ into two sets of $M$ corresponding regions, and penalize the discrepancy between the image $\mathbf{O}^{i}$ and the layer $\mathbf{G}^{i}$ on the average difference between the $j$-th ($j=1,...,M$) region and its neighboring regions.
Specifically, we perform average pooling on all elements in the divided regions, and calculate the differences between each pixel and its neighbors in the pooled regions for the image $\mathbf{O}^{i}$ and the layer $\mathbf{G}^{i}$, respectively.
Then we measure the discrepancy of spatial consistency between the image $\mathbf{O}^{i}$ and the layer $\mathbf{G}^{i}$ by mean squared error.
Formally, the ${\mathcal{L}_{ps}}$ is defined as: 
\begin{equation}
\setlength\abovedisplayskip{1pt}
\setlength\belowdisplayskip{1pt}
\label{loss:ps}
\begin{split}
{\mathcal{L}_{ps}} = \sum\limits_{i = 1}^n {{4^{(n - i)}}\frac{1}{M}{\sum\limits_{j = 1}^M {\sum\limits_{h \in \Omega (j)} \mathbf{(} {{\mathbf{Avg}(\mathbf{O}^i_h)} - {\mathbf{Avg}(\mathbf{O}^i_j)} } }}} -\\
{{\mathbf{Avg}(\mathbf{G}^i_h)} + \mathbf{Avg}(\mathbf{G}^i_j)} \mathbf{)}^2,
\end{split}
\end{equation}
where $M$ is the number of regions, $\Omega(j)$ denotes the regions centered at region $j$, $\mathbf{O}^{i}_j$ is the region $j$ of $i$-{th} layer image restored by our FCNet, $\mathbf{G}^{i}_j$ is the region $j$ of $i$-{th} pyramid layer decomposed from the ground-truth image $\mathbf{G}$, and $\mathbf{Avg}$ is the scalar averaging operation over all elements. 

\subsection{Implementation Details}
\label{sec:detail}
In our FCNet, we set $n=4$ in the Laplacian Pyramid decomposition.
The parameters of our FCNet are initialized by Kaiming-initialization~\cite{Kaiming} and optimized by Adam~\cite{AdamAM} with default parameters.
Our FCNet is trained in a total of 150 epochs.
The learning rate is initialized as $10^{-4}$ and decayed by a factor of $0.8$ for every $50$ epochs.
%
%
%
The $\mathcal{L}_{pr}$ is formulated in terms of summation over deviations on each pixels, while the $\mathcal{L}_{prs}$ is formulated on mean values.
Due to the magnitude difference between $\mathcal{L}_{pr}$ and $\mathcal{L}_{prs}$, we set $\lambda=4000$ to trade-off different loss terms, and will study the selection of $\lambda$ in ablation studies.
In the training stage, for each image sequence, we randomly select 1$\sim$10 images as input base sequence.
Except for Ablation settings, all testing results of our FCNet is evaluated from a single trained model on different tasks in~\S\ref{sec:experiment}.
Our FCNet is implemented in PyTorch~\cite{pytorch} and trained on a Titan RTX GPU with 24Gb memory.

\section{Experiment}
\label{sec:experiment}

\subsection{Dataset and Metric}\label{sec:dataset}
\noindent
\textbf{Dataset}.
We use the dataset in~\cite{afifi2021learning}, which contains 24,330 images.
As far as we know, it is the largest dataset with multi-exposed image sequences.
The images in~\cite{afifi2021learning} are rendered from MIT-Adobe FiveK dataset~\cite{bychkovsky2011learning} with 5 digital Exposure Values (EVs, \ie, -1.5, -1, +0, +1, +1.5) estimated by Adobe Camera Raw SDK.
This dataset is divided into a \texttt{training} set with 17,675 images, a \texttt{validation} set with 750 images, and a \texttt{test} set with 5,905 images.
As suggested in~\cite{gharbi2017deep, hu2018exposure, park2018distort}, we choose the images retouched by ``Expert C'' in~\cite{bychkovsky2011learning} as the ground-truths during the training and test stages.
\noindent
\textbf{Metric}.
We evaluate different methods by employing the widely used Peak Signal-to-Noise Ratio (PSNR) and Structural Similarity
Index (SSIM)~\cite{ssim} metrics.

\begin{figure}[!t]
\centering
\captionsetup[subfloat]{labelsep=none,format=plain,labelformat=empty,font={scriptsize,stretch=0.3}}
\subfloat[(1-a) Input Sequence]{
    \includegraphics[scale=0.232]{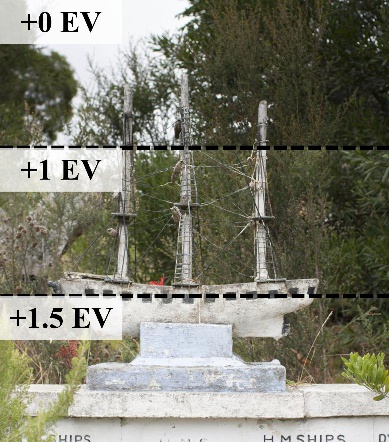}}
\subfloat[(1-b) Mertens09~\cite{mertens2009exposure}]{
    \includegraphics[scale=0.232]{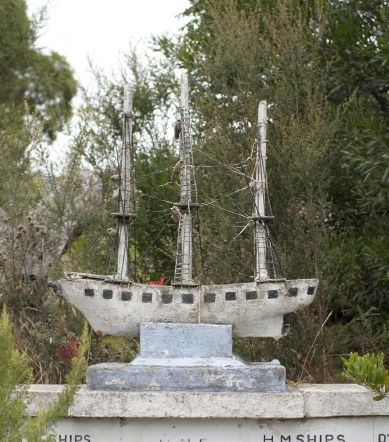}}
\subfloat[(1-c) MEF-GAN~\cite{xu2020GAnmef}]{
    \includegraphics[scale=0.232]{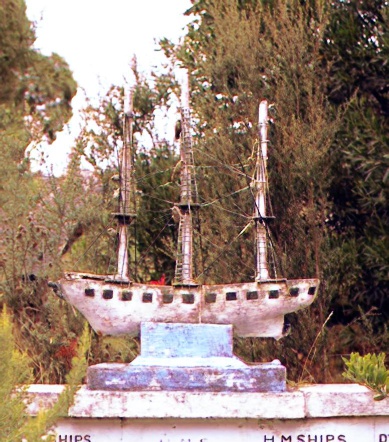}}
\subfloat[(1-d) MEF-Net~\cite{ma2019deep}]{
    \includegraphics[scale=0.232]{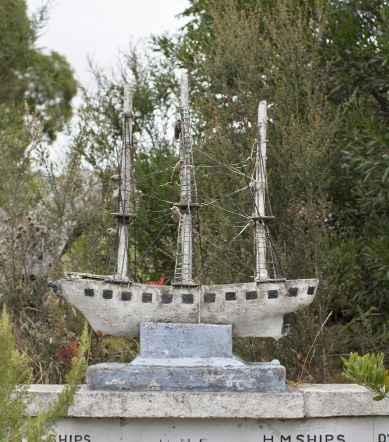}}
\\
\vspace{-0.2cm}
\subfloat[(1-e) Zero-DCE~\cite{guo2020zero}]{
    \includegraphics[scale=0.232]{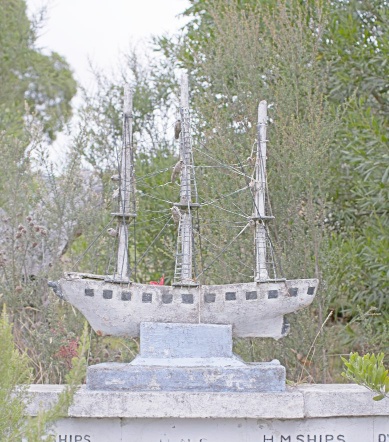}}
\subfloat[(1-f) MSEC~\cite{afifi2021learning}]{
    \includegraphics[scale=0.232]{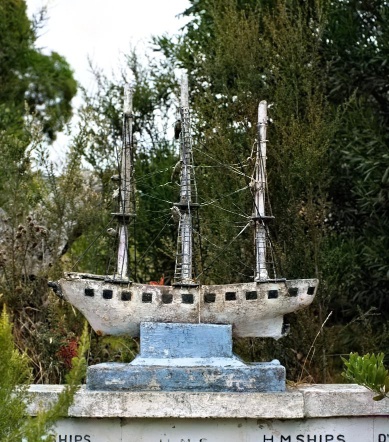}}
\subfloat[(1-g) FCNet (Ours)]{
    \includegraphics[scale=0.232]{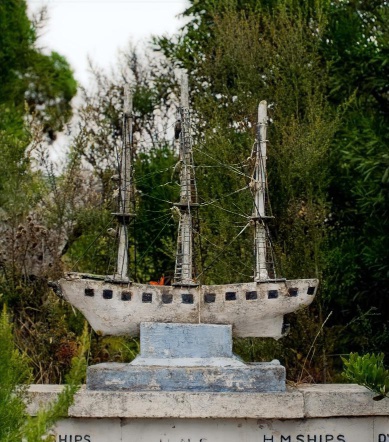}}
\subfloat[(1-h) Ground-Truth]{
    \includegraphics[scale=0.232]{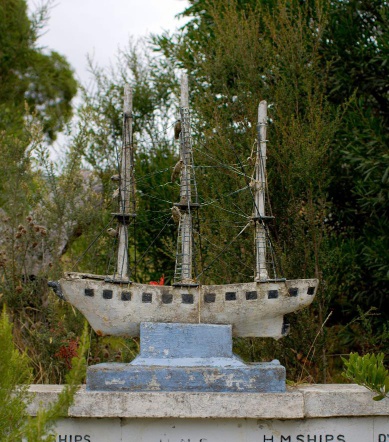}}
\\
\vspace{-0.2cm}
\subfloat[(2-a) Input Sequence]{
    \includegraphics[scale=0.20]{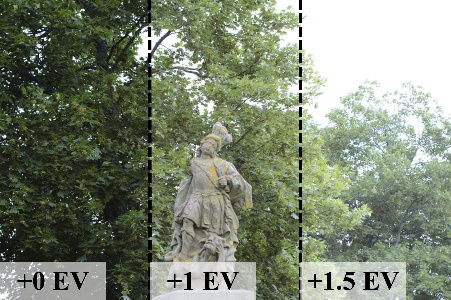}}
\subfloat[(2-b) Mertens09~\cite{mertens2009exposure}]{
    \includegraphics[scale=0.20]{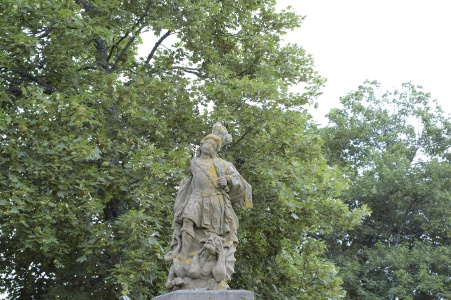}}
\subfloat[(2-c) MEF-GAN~\cite{xu2020GAnmef}]{
    \includegraphics[scale=0.20]{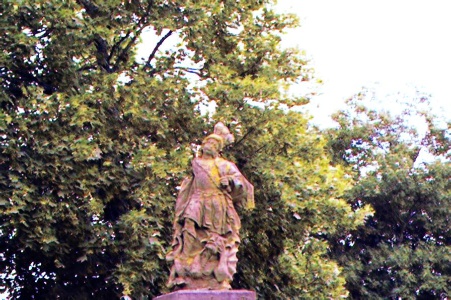}}
\subfloat[(2-d) MEF-Net~\cite{ma2019deep}]{
    \includegraphics[scale=0.20]{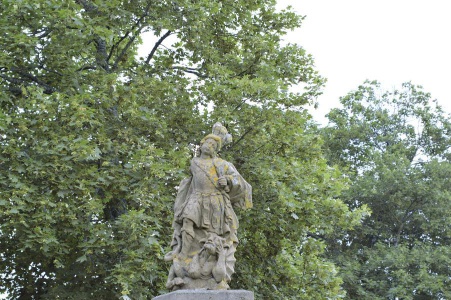}}
\\
\vspace{-0.2cm}
\subfloat[(2-e) Zero-DCE~\cite{guo2020zero}]{
    \includegraphics[scale=0.20]{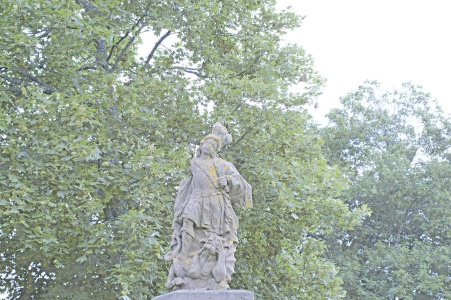}}
\subfloat[(2-f) MSEC~\cite{afifi2021learning}]{
    \includegraphics[scale=0.20]{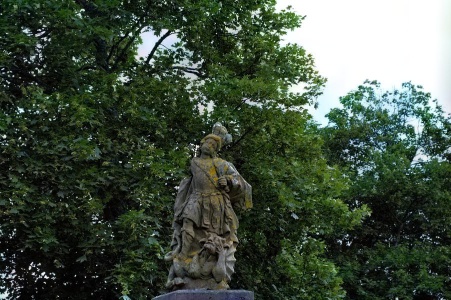}}
\subfloat[(2-g) FCNet (Ours)]{
    \includegraphics[scale=0.20]{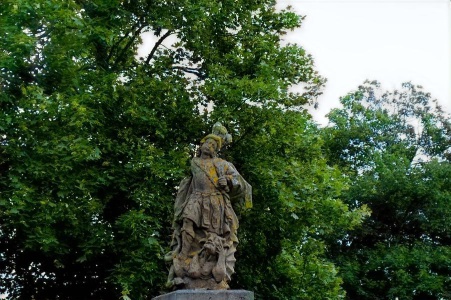}}
\subfloat[(2-h) Ground-Truth]{
    \includegraphics[scale=0.20]{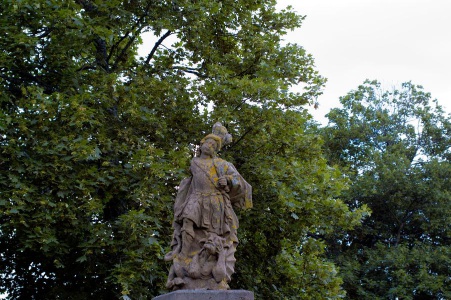}}
\vspace{-0.2cm}
\caption{\textbf{Qualitative comparison on an over-exposed image sequence} (with EVs of +0, +1, +1.5) by different MEF methods (Mertens09~\cite{mertens2009exposure}, MEF-GAN~\cite{xu2020GAnmef}, MEF-Net~\cite{ma2019deep}) and SEC methods (Zero-DCE~\cite{guo2020zero}, MSEC~\cite{afifi2021learning}).}
\vspace{-0.2cm}
\label{fig:overexposure comparison 1}
\end{figure}

\begin{figure}[!t]
\centering
\captionsetup[subfloat]{labelsep=none,format=plain,labelformat=empty,font={scriptsize,stretch=0.3}}
\subfloat[(a) Input Sequence]{
    \includegraphics[scale=0.285]{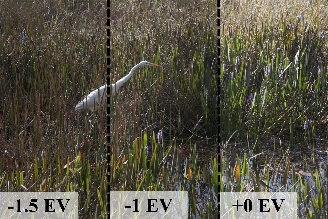}}
\subfloat[(b) Mertens09~\cite{mertens2009exposure}]{
    \includegraphics[scale=0.285]{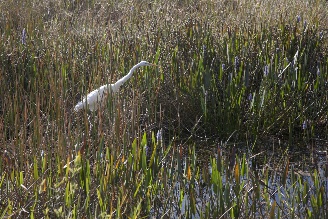}}
\subfloat[(c) MEF-GAN~\cite{xu2020GAnmef}]{
    \includegraphics[scale=0.285]{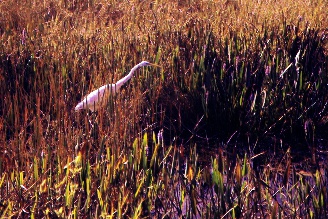}}
\subfloat[(d) MEF-Net~\cite{ma2019deep}]{
    \includegraphics[scale=0.285]{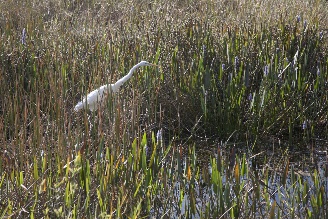}}
\\
\vspace{-0.2cm}
\subfloat[(e) Zero-DCE~\cite{guo2020zero}]{
    \includegraphics[scale=0.285]{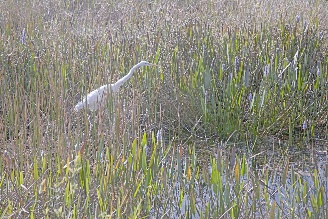}}
\subfloat[(f) MSEC~\cite{afifi2021learning}]{
    \includegraphics[scale=0.285]{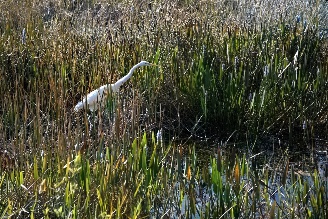}}
\subfloat[(g) FCNet (Ours)]{
    \includegraphics[scale=0.285]{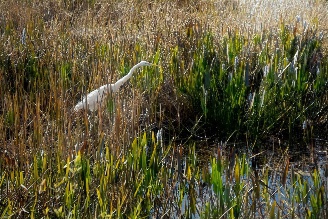}}
\subfloat[(h) Ground-Truth]{
    \includegraphics[scale=0.285]{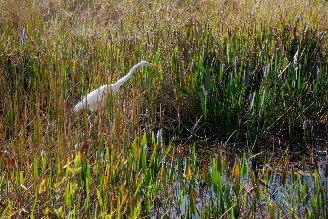}}
\\
\vspace{-0.2cm}
\caption{\textbf{Qualitative comparison on an under-exposed image sequence} (with EVs of +0, -1, -1.5) by different MEF methods.}
\label{fig:underexposure comparison 1}
\end{figure}

\subsection{Comparison on Multi-Exposure Fusion}\label{sec:MEF}
\noindent
\textbf{Experimental setting}.
We evaluate our FCNet on three fusion tasks: Under-Exposure Fusion (Under-EF) on three images rendered
with EVs of +0, -1, and -1.5, Over-Exposure Fusion (Over-EF) on three images rendered
with EVs of +0, +1, and +1.5, and Multi-Exposure Fusion (MEF) on all five images rendered
with EVs of -1, -1.5, +0, +1, and +1.5.

\noindent
\textbf{Comparison methods}.\ We compare our FCNet on the three fusion tasks with three MEF methods:\ Mertens09~\cite{mertens2009exposure}, MEF-GAN~\cite{xu2020GAnmef}, MEF-Net~\cite{ma2019deep}, and two single-exposure correction (SEC) methods:\ Zero-DCE~\cite{guo2020zero}, and MSEC~\cite{afifi2021learning}.
We evaluate these methods by their official implementations with default settings, except that we use the provided results by MSEC~\cite{afifi2021learning}.
These methods implementations are one single model for each method, and evaluated on different test tasks.
Since MEF-GAN only fuses two images, we run it for different two-exposure combinations and provide the results with the best PSNR results and corresponding SSIM results.
To perform Zero-DCE and MSEC on MEF, we use a heuristic approach that applies them on each image in a multi-exposure sequence, and averages the exposure-corrected images.
%

\begin{table}[]
\caption{\textbf{PSNR and SSIM~\cite{ssim} results of different methods on Multi-Exposure Fusion (MEF)}. The tasks are Under-Exposure Fusion (Under-EF, with EVs of +0, -1, -1.5), Over-Exposure Fusion (Over-EF, with EVs of +0, +1, +1.5), and All MEF (with all 5 EVs).
The best and second best results are highlighted in \textcolor{red}{\textbf{red}} and \textcolor{blue}{\textbf{blue}}, respectively.
}
\vspace{-0.8cm}
\begin{center}
\resizebox{0.8\textwidth}{!}{%
\begin{tabular}{|l|rr|rr|rr|}
\hline
\multirow{2}{*}{\diagbox{Method}{Setting}} & \multicolumn{2}{c|}{\begin{tabular}[c]{@{}l@{}}Under-EF\end{tabular}} & \multicolumn{2}{c|}{\begin{tabular}[c]{@{}l@{}}Over-EF \end{tabular}} & \multicolumn{2}{c|}{\begin{tabular}[c]{@{}l@{}}All MEF\end{tabular}} \\ \cline{2-7} 
& PSNR& SSIM& PSNR& SSIM& PSNR& SSIM\\ \hline
Mertens09~\cite{mertens2009exposure} &19.00  & 0.813   &17.39  & 0.804& 20.41 & \textcolor{blue}{\textbf{0.838}}\\
MEF-GAN~\cite{xu2020GAnmef} & 15.40 & 0.517 & 13.38 & 0.550 & 17.99 & 0.581\\
MEF-Net~\cite{ma2019deep} & 19.65& 0.825 & 17.85 & 0.803 &18.93 & 0.818\\ 
\hline
Zero-DCE~\cite{guo2020zero}& 13.92  & 0.717 & 10.39 & 0.669  &12.12 & 0.670\\
MSEC~\cite{afifi2021learning}& \textcolor{red}{\textbf{20.76}}  & \textcolor{red}{\textbf{0.829}} & \textcolor{blue}{\textbf{20.22}} & \textcolor{blue}{\textbf{0.831}}  & \textcolor{blue}{\textbf{20.66}} & 0.832\\
\hline
FCNet (Ours)  & \textcolor{blue}{\textbf{20.14}} & \textcolor{blue}{\textbf{0.816}} & \textcolor{red}{\textbf{20.41}} & \textcolor{red}{\textbf{0.839}} & \textcolor{red}{\textbf{20.81}} & \textcolor{red}{\textbf{0.847}} \\
\hline
\end{tabular}%
}
\end{center}
\vspace{-0.8cm}
\label{mef compare}
\end{table}

\noindent
\textbf{Quantitative results} on PSNR and SSIM by different methods are listed in Table~\ref{mef compare}.
%
%
One can see that our FCNet achieves consistently comparable or better results than the other
methods~\cite{ma2019deep,mertens2009exposure,xu2020GAnmef} on the Over-EF and MEF tasks, and comparable results on the Under-EF task. 
The MEF methods~\cite{ma2019deep,mertens2009exposure,xu2020GAnmef} can hardly achieve promising fusion results on the Under-EF or Over-EF tasks, which may be attributed to the lack of correction capability.
By simply averaging the corrected images, MSEC~\cite{afifi2021learning} achieves competing PSNR and SSIM results.
By integrating fusion and correction into a single framework, our FCNet is able to achieve robust results upon all three fusion tasks.

\noindent
\textbf{Visual quality}.
The results of MEF with all EVs by different methods are shown in Figure~\ref{fig:MEF comparison results 1}.
One can see that, when compared to the ground-truth, our FCNet well preserves the color and contrast information.
The Mertens09, MEF-Net, or MEF-GAN methods are prone to generate images with a certain contrast or color bias.
%
The result of Zero-DCE is over-exposed, while the result of MSEC presents a clear color shift. 

We also show the results on Over-EF in Figure~\ref{fig:overexposure comparison 1}.
We observe that the result of our FCNet is closer to the ground-truth than the other methods on the aspect of visual quality.
Due to a lack of correction capability, the methods of ``Mertens09'' and MEF-Net produce over-exposed images.
Performing Over-EF for MSEC by simple averaging brings a clear color bias, \eg, the ship.
As a low-light enhancer, Zero-DCE cannot obtain well-exposed results upon over-exposed images.
MEF-GAN suffers from clear color deviation on the Over-EF task, since it is trained on pairs of under- and over-exposed images.
%
Besides, as shown in Figure~\ref{fig:underexposure comparison 1}, on the Under-EF task, the image contrast and color saturation of the image produced by our FCNet are the closest to that of ground-truth over the other competitors.

\begin{figure}[!t]
\centering
\captionsetup[subfloat]{labelsep=none,format=plain,labelformat=empty,font={scriptsize,stretch=0.3}}
\subfloat[(a) Input Image]{
    \includegraphics[scale=0.285]{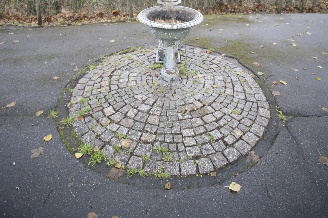}}
\subfloat[(b) LIME~\cite{guo2016lime}]{
    \includegraphics[scale=0.285]{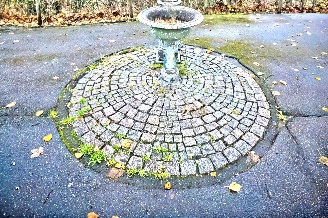}}
\subfloat[(c) WVM~\cite{fu2016weighted}]{
    \includegraphics[scale=0.285]{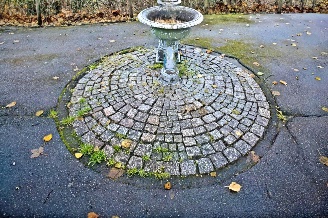}}
\subfloat[(d) EnlightenGAN~\cite{jiang2021enlightengan}]{
    \includegraphics[scale=0.285]{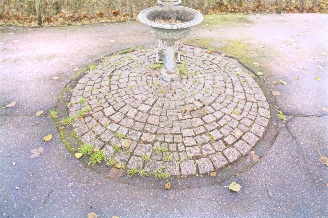}}
\\
\vspace{-0.2cm}
\subfloat[(e) Zero-DCE~\cite{guo2020zero}]{
    \includegraphics[scale=0.285]{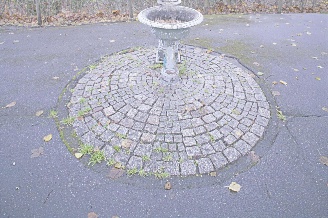}}
\subfloat[(f) MSEC~\cite{afifi2021learning}]{
    \includegraphics[scale=0.285]{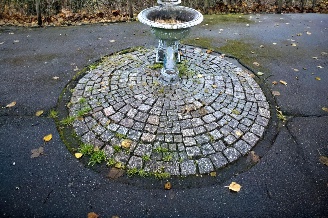}}
\subfloat[(g) FCNet (Ours)]{
    \includegraphics[scale=0.285]{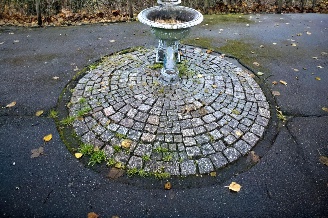}}
\subfloat[(h) Ground-Truth]{
    \includegraphics[scale=0.285]{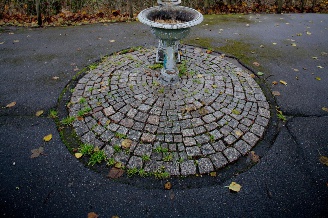}}
\\
\vspace{-0.2cm}
\caption{\textbf{Qualitative comparison on over-exposure correction task} by different single exposure correction methods.}
\vspace{-0.4cm}
\label{fig:SEC comparison results 1}
\end{figure}

\begin{figure}[!t]
\centering
\captionsetup[subfloat]{labelsep=none,format=plain,labelformat=empty,font={scriptsize,stretch=0.3}}
\subfloat[(a) Input Image]{
    \includegraphics[scale=0.285]{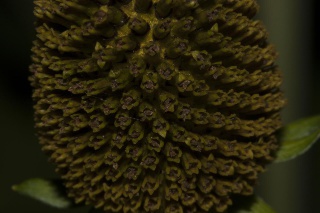}}
\subfloat[(b) LIME~\cite{guo2016lime}]{
    \includegraphics[scale=0.285]{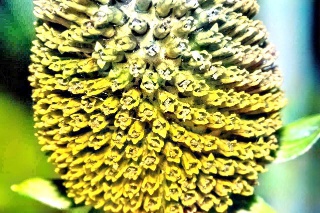}}
\subfloat[(c) WVM~\cite{fu2016weighted}]{
    \includegraphics[scale=0.285]{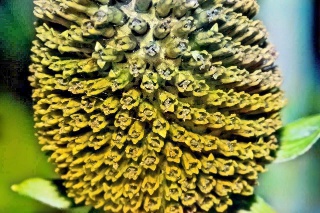}}
\subfloat[(d) EnlightenGAN~\cite{jiang2021enlightengan}]{
    \includegraphics[scale=0.285]{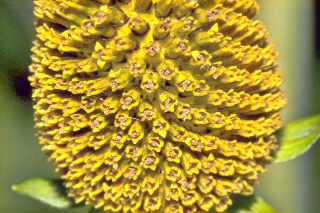}}
\\
\vspace{-0.2cm}
\subfloat[(e) Zero-DCE~\cite{guo2020zero}]{
    \includegraphics[scale=0.285]{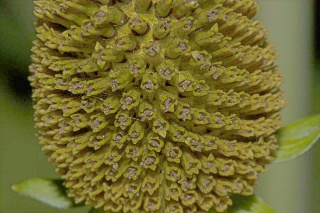}}
\subfloat[(f) MSEC~\cite{afifi2021learning}]{
    \includegraphics[scale=0.285]{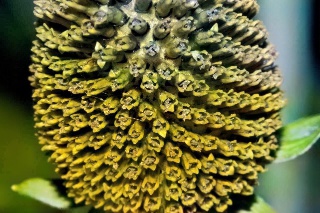}}
\subfloat[(g) FCNet (Ours)]{
    \includegraphics[scale=0.285]{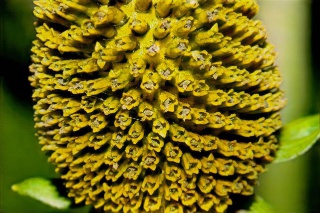}}
\subfloat[(h) Ground-Truth]{
    \includegraphics[scale=0.285]{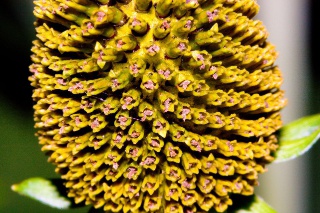}}
\\
\vspace{-0.2cm}
\caption{\textbf{Qualitative comparison on low-light enhancement task by different single exposure correction methods}.}
\vspace{-0.4cm}
\label{fig:SEC comparison results 2}
\end{figure}

\subsection{Comparison on Single Exposure Correction}\label{sec:SE}

\noindent
\textbf{Comparison methods}. Our FCNet is able to tackle single-exposure correction (SEC). Here we compare with 5 state-of-the-art SEC methods: WVM~\cite{fu2016weighted}, LIME~\cite{guo2016lime}, EnlightenGAN~\cite{jiang2021enlightengan}, Zero-DCE~\cite{guo2020zero}, and MSEC~\cite{afifi2021learning}.

\noindent
\textbf{Quantitative comparisons}.
In Table~\ref{single compare}, we observe that our FCNet arrives at comparable PSNR and SSIM results with MSEC~\cite{afifi2021learning} and much better results than the others.
Our FCNet is slightly better than the MSEC on the over-exposure correction tasks, while is slightly inferior to the MSEC on under-exposure correction. 
These results validate the flexibility and effectiveness of our FCNet on SEC.

\noindent
\textbf{Visual quality}.
As shown in Figure~\ref{fig:SEC comparison results 1}, similar to MSEC~\cite{afifi2021learning}, our FCNet achieves visually favorable results on correcting over-exposed images.
By contrast, the methods of LIME and WVM produce over-bright images, while EnlightenGAN and Zero-DCE have obvious flaws such as whitening and artifacts.
For under-exposed correction (low-light enhancement), as shown in Figure~\ref{fig:SEC comparison results 2}, MSEC, LIME and WVM lead to some over-exposure artifacts (\eg, the top of the image), while EnlightenGAN and Zero-DCE generate unsatisfactory visual results in terms of both image contrast and naturalness.
On the contrary, our FCNet generates proper results with reasonable image
contrast and vivid color.
This demonstrates that our FCNet is also very effective on the SEC task.


\begin{table}[!h]
\caption{ \textbf{Average PSNR and SSIM~\cite{ssim} results of different methods on Single Exposure Correction (SEC)}. We compare different methods on properly exposed and underexposed images (+0, -1, -1.5 EVs) correction, overexposed images (+1, +1.5 EVs) correction, and correcting all five images. The best and second best results are highlighted in \textcolor{red}{\textbf{red}} and \textcolor{blue}{\textbf{blue}}, respectively.
}
\vspace{-0.8cm}
\begin{center}
\resizebox{0.8\textwidth}{!}{%
\begin{tabular}{|r|rr|rr|rr|}
\hline
\multirow{2}{*}{\diagbox{Method}{Exposure}} & \multicolumn{2}{c|}{\begin{tabular}[c]{@{}l@{}}+0, -1, -1.5 EVs\end{tabular}} & \multicolumn{2}{c|}{\begin{tabular}[c]{@{}l@{}}+1, +1.5 EVs\end{tabular}} & \multicolumn{2}{c|}{\begin{tabular}[c]{@{}l@{}}All\end{tabular}} \\ \cline{2-7} 
& PSNR& SSIM& PSNR& SSIM& PSNR& SSIM\\ \hline
LIME~\cite{guo2016lime} &12.01  & 0.630  & 11.62  & 0.619 & 11.86 & 0.626\\
WVM~\cite{fu2016weighted} & 17.01 & 0.741  & 16.30&0.735& 16.72 & 0.739\\
EnlightenGAN~\cite{jiang2021enlightengan} & 12.38& 0.696 & 8.91  & 0.629 &10.99 & 0.669\\
Zero-DCE~\cite{guo2020zero}& 13.95  & 0.699 & 9.33  & 0.620&12.10 & 0.668\\
MSEC~\cite{afifi2021learning}& \textcolor{red}{\textbf{20.44}}  & \textcolor{red}{\textbf{0.810}} & \textcolor{blue}{\textbf{19.53}}  & \textcolor{blue}{\textbf{0.813}} &\textcolor{red}{\textbf{20.08}} & \textcolor{red}{\textbf{0.811}}\\
FCNet (Ours) & \textcolor{blue}{\textbf{19.69}}  & \textcolor{blue}{\textbf{0.788}} & \textcolor{red}{\textbf{19.63}} & \textcolor{red}{\textbf{0.816}} &\textcolor{blue}{\textbf{19.67}} & \textcolor{blue}{\textbf{0.799}}\\
\hline
\end{tabular}%
}
\end{center}
\vspace{-0.8cm}
\label{single compare}
\end{table}

\subsection{Ablation Study}\label{sec:ablation}
We now conduct detailed examinations of our FCNet on the SEC and MEF tasks to assess:
1) the effectiveness of gradually decreasing the size of the proposed FC block during the image reconstruction progress;
%
%
2) the necessity of combining Fusion and Correction blocks to process arbitrary number of input frames;
3) the order of Fusion and Correction blocks;
4) the influence of loss functions to our FCNet;
5) the choice of $\lambda$ to trade off the loss function $\mathcal{L}_{pr}$ and $\mathcal{L}_{ps}$.

\begin{figure}[!t]
\centering
\captionsetup[subfloat]{labelsep=none,format=plain,labelformat=empty,font={scriptsize,stretch=0.3}}
\subfloat[(a) Input Sequence]{
    \includegraphics[scale=0.18]{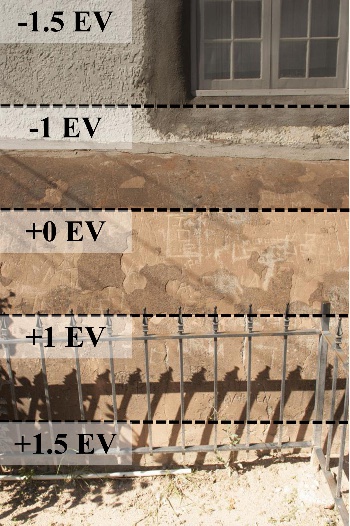}}
\subfloat[(b) $\mathcal{L}_{r}$]{
    \includegraphics[scale=0.18]{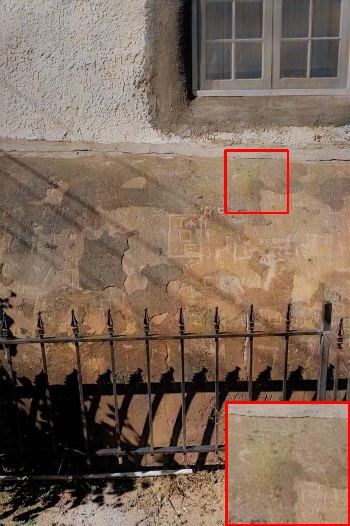}}
\subfloat[(c) $\mathcal{L}_{pr}$]{
    \includegraphics[scale=0.18]{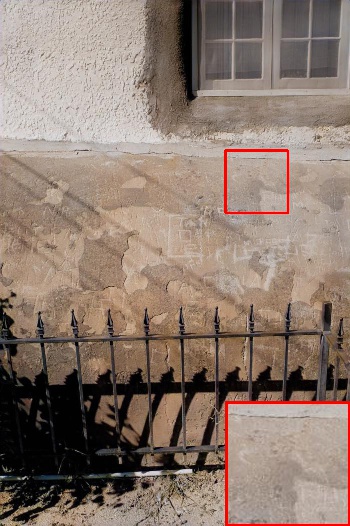}}
\subfloat[(d) $\mathcal{L}_{pr}+\mathcal{L}_{s}$]{
    \includegraphics[scale=0.18]{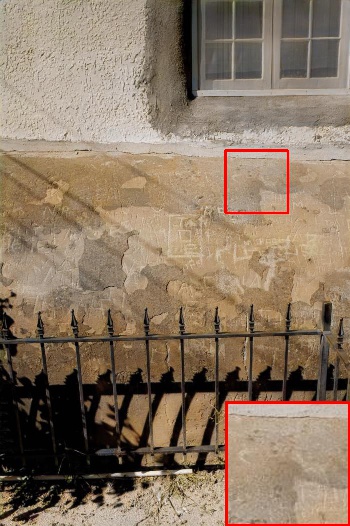}}
\subfloat[(e) $\mathcal{L}_{pr}+\mathcal{L}_{ps}$]{
    \includegraphics[scale=0.18]{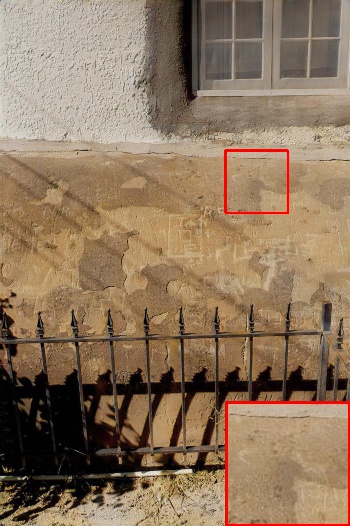}}
\subfloat[(f) Ground-Truth]{
    \includegraphics[scale=0.18]{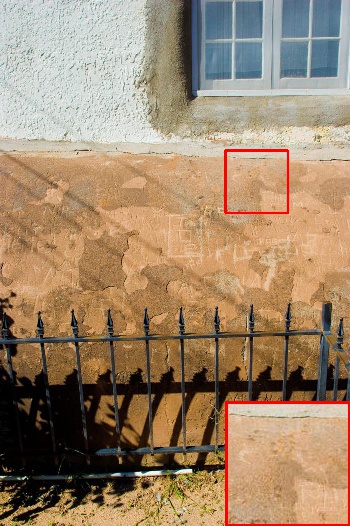}}
\vspace{-0.2cm}
\caption{\textbf{Ablation study of the contribution of each loss} (pyramid reconstruction loss $\mathcal{L}_{pr}$ and pyramid space consistency loss $\mathcal{L}_{ps}$) with EVs of -1.5, -1, +0, +1, +1.5. Also, in order to prove the necessity of using loss function to constrain the output of each layer of model, we compared the effect of loss functions which only applying in the last layer output of the model (reconstruction loss $\mathcal{L}_{r}$ and space consistency loss $\mathcal{L}_{s}$). \textcolor{red}{Red} boxes indicate the obvious differences and amplified details.}
\vspace{-0.4cm}
\label{figure:loss function}
\end{figure}

\noindent
\textbf{1) The effectiveness of gradually decreasing the size of FC block during the progress of the image reconstruction.}
As most of the illumination and color information of the image sequence is mainly extracted at the bottom LP levels, it is efficient to use larger networks to fuse and correct the bottom levels with lower resolutions and smaller networks to process the top LP levels with higher resolutions~\cite{liang2021high}, as introduced in \S\ref{sec:FC}.
To validate the effectiveness of this design for our FCNet, we compare three different network design schemes, which are described as follows.

\noindent
\textbf{Model 1}.
As the basis for comparison, this model applies the smallest Fusion block and Correction block in each LP level.
The parameter $m$ of each fusion block (Figure~\ref{fig: fusion_block}) is set as $0$, and each correction block (Figure~\ref{fig: correction_block}) is a $3$-layer UNet-like network with $16$ channels in the first convolutional layer of the encoder.
We denote this model as ``Small$\rightarrow$Small''.

\noindent
\textbf{Model 2}.
In this model, the size of FC block gradually increases with the progress of the image reconstruction.
For different LP levels from bottom to top, the parameter $m$ of four Fusion blocks are set as $0,1,2,3$, respectively.
The first and second Correction blocks are 3-layer UNet-like networks, with $16$ channels in the first convolutional layer of the encoder.
The third Correction block is a 4-layer UNet-like network and has $16$ channels in the first convolutional layer of the encoder.
The final ($4$-th) Correction block is a 4-layer UNet-like network, but with $24$ channels in the first convolutional layer of the encoder.
We denote this model as ``Small$\rightarrow$Large''.

\noindent
\textbf{Model 3}.
Contrary to the \textbf{Model 1}, this model applies the largest Fusion block and Correction block to each LP level.
The parameter $m$ of each Fusion block is set as $3$.
All the Correction blocks are 4-layers UNet-like networks, with $24$ channels in the first convolutional layer of the encoder.
We denote this model as ``Large$\rightarrow$Large''.

\noindent
\textbf{Model 4}. This is our FCNet, in which the FC blocks are introduced in \S\ref{sec:FC}.
We denote this model as ``Large$\rightarrow$Small''.

As shown in Table~\ref{table:modelsize}, our FCNet (Model 4, ``Large$\rightarrow$Small'') with decreasing amounts of model size during the image reconstruction progress obtains better PSNR and SSIM results on MEF, similar results on Under-EF, while comparable results on SEC and Over-EF, when compared with the Model 3 which employs the largest Fusion block and Correction block in each LP level.
At the same time, the Model 1 (``Small$\rightarrow$Small'') with the smallest Fusion block and Correction block has the fewest learnable parameters and FLOPs, but cannot well handle the tasks.
It is worth noting that the FLOPs of our FCNet and Model 1 are similar, but our FCNet achieves better performance than Model 1 on both SEC and MEF tasks.
Thus it is reasonable and effective to allocate more parameters and computational costs to the bottom LP levels, which contain more visual attributes than the top LP levels under the LP decomposition framework.

\begin{table*}[]
\caption{\textbf{PSNR and SSIM results of our FCNet with difference designs for the FC block} on the Single-Exposure Correction (SEC) and Multi-Exposure Fusion (MEF) tasks.
The best results are highlighted in \textbf{bold}.
We also compare the number of learnable parameters (Params.) and FLOPs upon processing five $512 \times 512$ multi-exposed images, in our FCNet.}
\vspace{-0.8cm}
\begin{center}

\resizebox{\textwidth}{!}{%
\begin{tabular}{|c|cc|cc|cc|cc|cc|}
\hline
\multirow{2}{*}{\diagbox {Model}{Task}} & \multicolumn{2}{c|}{SEC} & \multicolumn{2}{c|}{Under-EF} & \multicolumn{2}{c|}{Over-EF} & \multicolumn{2}{c|}{MEF} & \multicolumn{2}{c|}{Network Complexity} \\ \cline{2-11} 
        & PSNR    & SSIM   & PSNR    & SSIM   & PSNR    & SSIM   & PSNR    & SSIM & Params. $(\times M)$ &  FLOPs $(\times G)$  \\ \hline
Small$\rightarrow$Small & 18.04& 0.7780 &18.65&0.7649&19.01&0.7816& 19.75 & 0.8228&0.530&12.16\\
Small$\rightarrow$Large &18.56&0.7816&18.86&0.7608&19.27&0.7878&20.11 &0.8379&2.055&34.18\\
Large$\rightarrow$Large &19.27&\textbf{0.8025}&\textbf{20.51}&\textbf{0.8271}&20.38&\textbf{0.8411}&20.42&0.8444&4.904&39.78\\
Large$\rightarrow$Small & \textbf{19.67} & 0.7991& 20.14 & 0.8160& \textbf{20.41} & 0.8393 & \textbf{20.81} & \textbf{0.8465}&2.055&12.90\\ \hline
\end{tabular}%
}
\end{center}
\vspace{-0.4cm}
\label{table:modelsize}
\end{table*}

\noindent
\textbf{2) The necessity of combining Fusion block and Correction block to process input image sequence with arbitrary length}.
To study this problem, we develop two variants of our FCNet by only using Fusion blocks (``Fusion Only'') or only using Correction blocks (``Correction Only'').
The comparison results shown in Table~\ref{fusion only or correction only} demonstrate that, the variant ``Fusion Only'' well handles the MEF task but fails on the SEC task.
The variant ``Correction Only'' can handle both the SEC and MEF tasks.
Note that for MEF, it performs correction image-by-image and simply averages the corrected images.
By combining Fusion and Correction blocks, our FCNet obtains robust PSNR/SSIM results on both SEC and MEF.

\begin{table}[]
\begin{center}
\caption{\textbf{Results of our FCNet with only fusion or correction blocks} on Single-Exposure Correction (SEC) and Multi-Exposure Fusion (MEF) tasks.
Note that the ``Fusion Only'' model does not change the input image on SEC.
}
\vspace{-0.3cm}
\resizebox{0.6\textwidth}{!}{%
\begin{tabular}{|r|cc|cc|}
\hline
\multirow{2}{*}{\diagbox {Model}{Task}}      & \multicolumn{2}{c|}{\begin{tabular}[c]{@{}l@{}}SEC\end{tabular}} & \multicolumn{2}{c|}{\begin{tabular}[c]{@{}l@{}}MEF\end{tabular}} \\ \cline{2-5} 
& PSNR& SSIM& PSNR & SSIM \\ \hline
Fusion Only & 16.40& 0.7399& 21.67&0.8364\\
Correction Only & 19.69& 0.8202 & 20.05& 0.8386\\
Our FCNet & 19.67& 0.7991 & 20.81& 0.8465\\ \hline
\end{tabular}
}
\label{fusion only or correction only}
\end{center}
\vspace{-0.4cm}
\end{table}

\noindent
\textbf{3) The order between Fusion and Correction blocks}.
In our FCNet, we implement the Fusion block before the Correction block in each level of LP decomposition.
If we change the order of these two blocks in all 4 LP levels, the resulting computational costs and memory consumption is prohibitive on our RTX GPU with 24G memory.
Thus, here we only change the order in the first FC Block of our FCNet for comparisons.
As shown in Table~\ref{table:order of fusion and correction}, the results of ``first Correction then Fusion'' (Correction$\rightarrow$ Fusion) are slightly inferior to our original FCNet (Fusion$\rightarrow$ Correction).
Since it is prone to introduce unpleasing artifacts when correcting challenging over- or under-exposed images, applying fusion first in our FCNet would generate a nearly well exposed image and thereby reduce the correction difficulty.
In addition, compared with correcting a single fusion image, correcting a multi-exposure image sequence before fusion needs more computational costs and larger GPU memory.
Therefore, we finally adopt the idea of ``first Fusion then Correction''.
\begin{table}[t]
\caption{\textbf{Results of our FCNet with different orders of fusion and correction blocks} on Single-Exposure Correction (SEC) and Multi-Exposure Fusion (MEF) tasks.}
\vspace{-0.7cm}
\begin{center}
\resizebox{0.6\textwidth}{!}{%
\begin{tabular}{|c|cc|cc|}
\hline
\multirow{2}{*}{\diagbox {Model}{Task}}& \multicolumn{2}{c|}{\begin{tabular}[c]{@{}l@{}}SEC\end{tabular}} & \multicolumn{2}{c|}{\begin{tabular}[c]{@{}l@{}}MEF\end{tabular}} \\ \cline{2-5} 
& PSNR& SSIM& PSNR & SSIM \\ \hline
Correction$\rightarrow$ Fusion& 19.31& 0.794& 20.71&0.838\\
Fusion$\rightarrow$ Correction& 19.67& 0.799& 20.81& 0.847\\ \hline
\end{tabular}
}
\end{center}
\vspace{-0.4cm}
\label{table:order of fusion and correction}
\end{table}


\noindent
\textbf{4) The contribution of each loss term to our FCNet on SEC and MEF}.
To study the effect of different reconstruction and spatial loss functions to our FCNet on constraining the output of each LP level, we evaluate our FCNet by employing these loss functions only on the final output image.
We denote ${\mathcal{L}_{r}}$ as the top-level pyramid reconstruction loss function only on the final output, and denote the ${\mathcal{L}_{s}}$ as the spatial loss function only on the final output.
The ${\mathcal{L}_{r}}$ and ${\mathcal{L}_{s}}$ can be viewed as general forms of the ${\mathcal{L}_{pr}}$ in Eq.~(\ref{loss:pr}) and ${\mathcal{L}_{ps}}$ in Eq.~(\ref{loss:ps}), respectively with $i=1$.
As shown in Figure \ref{figure:loss function},
when our FCNet is trained only with $\mathcal{L}_{r}$, it would produce some unnatural color spots.
Our FCNet trained with the full pyramid reconstruction loss $\mathcal{L}_{pr}$ alleviates this problem, but produces results with low contrasts.
Besides, our FCNet trained with both $\mathcal{L}_{pr}$ and $\mathcal{L}_{s}$ produces brighter results.
By combining the full pyramid reconstruction loss $\mathcal{L}_{pr}$ and our pyramid spatial consistency loss $\mathcal{L}_{ps}$, our FCNet excludes the influence of artifacts to produce well-exposed images.
%
%
Table~\ref{table:loss function} also shows that  our FCNet trained with $\mathcal{L}_{pr}$ and $\mathcal{L}_{ps}$ attains the best results.

\begin{table}
\caption{\textbf{Results of our FCNet with different loss functions} on Single-Exposure Correction (SEC) and Multi-Exposure Fusion (MEF).\ The best results are shown in \textbf{bold}.}
\vspace{-0.6cm}
\begin{center}
\resizebox{0.55\textwidth}{!}{%
\begin{tabular}{|l|cc|cc|}
\hline
\multirow{2}{*}{\diagbox {Loss}{Task}} & \multicolumn{2}{c|}{\begin{tabular}[c]{@{}l@{}}SEC\end{tabular}} & \multicolumn{2}{c|}{\begin{tabular}[c]{@{}l@{}}MEF\end{tabular}} \\ \cline{2-5} & PSNR & SSIM & PSNR  & SSIM      \\ \hline
$\mathcal{L}_{r}$  & 19.14 & 0.7848& 20.14& 0.8208 \\
$\mathcal{L}_{pr}$ & 19.27 & \textbf{0.8042}& 20.13 & 0.8420 \\
$\mathcal{L}_{pr}+\mathcal{L}_{s}$ & 18.73 & 0.7822 & 19.21& 0.8100 \\
$\mathcal{L}_{pr}+\mathcal{L}_{ps}$  & \textbf{19.67} & 0.7991& \textbf{20.81}& \textbf{0.8465}\\ \hline
\end{tabular}
}
\label{table:loss function}
\end{center}
\vspace{-0.6cm}
\end{table}

\begin{table}[t]
\begin{center}
\caption{\textbf{PSNR and SSIM results by our FCNet with different $\lambda$ on the tasks of SEC and MEF (with all 5 EVs)}. The best and second best results are highlighted in \textcolor{red}{\textbf{red}} and \textcolor{blue}{\textbf{blue}}, respectively.
}
\vspace{-3mm}
\resizebox{0.45\textwidth}{!}{%
\begin{tabular}{|r|cc|cc|}
\hline
\multirow{2}{*}{\diagbox {$\lambda$}{Task}} & \multicolumn{2}{c|}{\begin{tabular}[c]{@{}l@{}}SEC\end{tabular}} & \multicolumn{2}{c|}{\begin{tabular}[c]{@{}l@{}}MEF\end{tabular}} \\ \cline{2-5} & PSNR & SSIM & PSNR  & SSIM      \\ \hline
1 & 18.64 &0.797 &19.62&0.837\\
10 & 19.32 &0.797 &\textcolor{blue}{\textbf{20.63}}&0.838\\
100 & \textcolor{blue}{\textbf{19.62}} &0.798 &20.37&0.831\\
1000  & 18.56 & 0.767 &19.27&0.804 \\
2000 &19.37 & \textcolor{blue}{\textbf{0.802}} &20.33&0.839\\
3000 &19.40 & 0.786 &\textcolor{blue}{\textbf{20.63}}&0.829\\ 
4000 & \textcolor{red}{\textbf{19.67}} & 0.799 &\textcolor{red}{\textbf{20.81}}&\textcolor{red}{\textbf{0.847}} \\
5000  &19.50 & \textcolor{red}{\textbf{0.806}} &20.61&\textcolor{blue}{\textbf{0.845}}\\
\hline
\end{tabular}

}
\vspace{-8mm}
\label{lambda ablation}
\end{center}
\end{table}

\noindent
\textbf{5) The choice of $\lambda$ to trade off $\mathcal{L}_{pr}$ and $\mathcal{L}_{ps}$.
}
The two loss functions $\mathcal{L}_{pr}$ and $\mathcal{L}_{ps}$ have different magnitudes: the $\mathcal{L}_{pr}$ is formulated in terms of summation over deviations on each pixels,  while the $\mathcal{L}_{ps}$ is formulated on mean values. Thus, it is reasonable to set a large $\lambda$ to balance the magnitude difference. The ablation studies on $\lambda$ is shown in Table~\ref{lambda ablation}. One can see that our FCNet is robust to the selection of $\lambda$ and achieves the best results in terms of PSNR and SSIM when $\lambda=4,000$.

\subsection{Comparison on Model Complexity and Speed}
\label{sec:comput&time}
%
%
In Table~\ref{table:time}, we compare the parameters amounts, FLOPs and running time of different methods on the SEC and MEF tasks.
On both tasks, we provide the results of different methods in correcting images of sizes $256\times256$, $512\times512$, or $1024\times1024$.
Since MSEC is originally implemented in Matlab, we also provide its running time  by using an unofficial PyTorch code\tablefootnote{https://github.com/LZ-CH/Exposure\_Correction-pytorch.}.
Note that we compute the FLOPs and running time of MEF-GAN for fusing two images since it can only process two images with different exposures.
The running time of different methods is evaluated on a machine with an Intel i9 10920X CPU and a Titan RTX GPU. We observe that our FCNet is faster than the comparison methods, except for Zero-DCE on SEC.
By employing the LP decomposition, our FCNet assigns minor computational costs on high-resolution top LP levels, and thus its running time is relatively robust to the increasing of image resolutions over the other methods.

\begin{table}[t]
\caption{\textbf{Comparison of parameter amounts, computational costs, and running time by different methods} on SEC and MEF. On MEF, we compare the FLOPs (G) and running time (second) by processing five multi-exposed images in different resolutions. ``$^*$'': MEF-GAN can only fuse two images with different exposures.
}
\begin{center}
\vspace{-6mm}
\resizebox{0.8\textwidth}{!}{%
\begin{tabular}{|c|l|c|rc|rc|rc|}
\hline
\multirow{2}{*}{Task} &\multirow{2}{*}{Methods} & \multirow{2}{*}{Params $(\times M)$} & \multicolumn{2}{c|}{$256 \times 256$} & \multicolumn{2}{c|}{$512\times512$} & \multicolumn{2}{c|}{$1024\times1024$} \\ \cline{4-9} 
         &&  & FLOPs & Time & FLOPS & Time & FLOPs & Time \\ \hline
\multirow{7}{*}{SEC}&LIME & \multicolumn{1}{c|}{-}  &  \multicolumn{1}{c}{-}     &1.065&\multicolumn{1}{c}{-}&2.905 &\multicolumn{1}{c}{-}& 11.138\\
&WVM & \multicolumn{1}{c|}{-}  &  \multicolumn{1}{c}{-}     &0.420&\multicolumn{1}{c}{-}&2.509 &\multicolumn{1}{c}{-}& 9.888\\
&EnlightenGAN &8.637 &12.59&0.006& 65.78&0.017&263.12&0.073\\
&Zero-DCE &0.079& 5.20 & 0.002 & 20.76 & 0.006&83.05&0.014\\
&MSEC (Matlab)&7.015&4.58&0.187&18.34&0.196&73.35&0.456\\
&MSEC (Pytorch)&7.015&4.58&0.014&18.34&0.025&73.35&0.078\\
&FCNet (Ours)   & 2.055&3.16&0.014&12.65&0.014& 50.61&0.023\\ \hline
\multirow{4}{*}{MEF}&Mertens09  & \multicolumn{1}{c|}{-}  &  \multicolumn{1}{c}{-}     &0.009&\multicolumn{1}{c}{-}&0.041 &\multicolumn{1}{c}{-}& 0.262\\
&MEF-GAN*  & 0.488  &  61.87     &0.950&247.49&2.348 &989.95& 8.474\\
&MEF-Net  &0.026  &8.62&0.008& 34.48 &0.030&137.90&0.121\\
&FCNet (Ours)   & 2.055&3.22&0.014&12.90&0.016&51.58&0.033\\ 
\hline
\end{tabular}%
}
\end{center}
\vspace{-8mm}
\label{table:time}
\end{table}


\section{Conclusion}
\label{sec:conclusion}
In this paper, we proposed a Fusion-Correction Network (FCNet) to simultaneously tackle the Single-Exposure Correction (SEC) and Multi-Exposure Fusion (MEF) tasks in an integrated framework.
In our FCNet, we implemented Laplacian Pyramid (LP) decomposition to exploit multi-scale context information of natural images.
In each LP level, we processed the base image sequence by our Fusion and Correction blocks sequentially.
The Fusion is implemented before correction to reduce the computation costs.
The Fusion block is a light-weight network for efficiency consideration, while the Correction block is a UNet-like network to enhance the exposure of the fused image.
The corrected image is upsampled and composed with the high-frequency components in the next LP level.
By integrating of fusion and correction blocks, our FCNet is feasible to process an image sequence of arbitrary length (including one).
Experimental results on the benchmark dataset~\cite{afifi2021learning} demonstrated that, our FCNet not only achieves competitive or even better performance on both MEF and SEC tasks when compared to the corresponding state-of-the-arts methods, but also well resolves the Over-EF and Under-EF tasks that previous MEF methods usually fail at.




\section{Acknowledgement}
This work was supported in part by The National Natural Science Foundation of China (No. 62002176 and 62176068), and sponsored by CAAI-Huawei MindSpore Open Fund.

\bibliographystyle{elsarticle-num}
\bibliography{bibtex}

\end{document}